\documentclass[
reprint, 
prd, 
aps, 
nofootinbib,
showpacs, 
superscriptaddress]{revtex4-2}

\usepackage{graphicx}
\usepackage[usenames,dvipsnames]{xcolor}
\usepackage[utf8]{inputenc}
\usepackage[T1]{fontenc}
\usepackage{anyfontsize}
\usepackage{adjustbox}
\usepackage{placeins}
\usepackage{lipsum}
\usepackage{latexsym}
\usepackage{rotating}
\usepackage{multirow}
\usepackage{mathtools}
\usepackage{amsmath, amssymb, amsthm, amsfonts}
\usepackage{mathrsfs}
\usepackage{bm}
\usepackage{enumerate}
\usepackage{aas_macros}
\usepackage{soul}
\usepackage{natbib}
\usepackage{hyperref}
\usepackage{orcidlink}
\usepackage{verbatim}
\usepackage[normalem]{ulem}
\hypersetup{colorlinks=true,citecolor=NavyBlue,linkcolor=NavyBlue,urlcolor=NavyBlue}

\newcommand{\ICASU}{\affiliation{Illinois Center for Advanced Studies of the Universe, Department of Physics, University of Illinois at Urbana-Champaign, Urbana, Illinois 61801, USA}}
\newcommand{\CAPS}{\affiliation{Center for AstroPhysical Surveys, National Center for Supercomputing Applications, Urbana, Illinois 61801, USA}}
\newcommand{\LIGOLAB}{\affiliation{LIGO Laboratory and Kavli Institute for Astrophysics and Space Research,
Massachusetts Institute of Technology, 185 Albany Street, Cambridge, Massachusetts 02139, USA}}
\newcommand{\UIUCASTRO}{\affiliation{Department of Astronomy, University of Illinois at Urbana-Champaign, Urbana, Illinois 61801, USA}}

\begin{document}

\title{Neural post-Einsteinian framework for efficient theory-agnostic tests of general relativity with gravitational waves}

\author{Yiqi~Xie~\orcidlink{0000-0002-8172-577X}}
\ICASU
\author{Deep~Chatterjee~\orcidlink{0000-0003-0038-5468}}
\LIGOLAB
\author{Gautham~Narayan~\orcidlink{0000-0001-6022-0484}}
\CAPS\UIUCASTRO
\author{Nicol\'as~Yunes~\orcidlink{0000-0001-6147-1736}}
\ICASU

\date{\today}

\begin{abstract}
The parametrized post-Einsteinian (ppE) framework and its variants are widely used to probe gravity through gravitational-wave tests that apply to a large class of theories beyond general relativity.
However, the ppE framework is not truly theory-agnostic as it only captures certain types of deviations from general relativity: those that admit a post-Newtonian series representation in the inspiral of coalescing compact objects. 
Moreover, each type of deviation in the ppE framework has to be tested separately, making the whole process computationally inefficient and expensive, possibly obscuring the theoretical interpretation of potential deviations that could be detected in the future. 
We here present the neural post-Einsteinian (npE) framework, an extension of the ppE formalism that overcomes the above weaknesses using deep-learning neural networks. 
The core of the npE framework is a variational autoencoder that maps the discrete ppE theories into a continuous latent space in a well-organized manner. This design enables the npE framework to test many theories simultaneously and to select the theory that best describes the observation in a single parameter estimation run. The smooth extension of the ppE parametrization also allows for more general types of deviations to be searched for with the npE model.
We showcase the application of the new npE framework to future tests of general relativity with the fifth observing run of the LIGO-Virgo-KAGRA collaboration.
In particular, the npE framework is demonstrated to efficiently explore modifications to general relativity beyond what can be mapped by the ppE framework, including modifications coming from higher-order curvature corrections to the Einstein-Hilbert action at high post-Newtonian order, and dark-photon interactions in possibly hidden sectors of matter that do not admit a post-Newtonian representation.
\end{abstract}

\maketitle

\section{Introduction}
Although general relativity (GR) is currently our best theory of gravity, it has been challenged by theoretical issues and observational anomalies.
From the theoretical perspective, for example, GR unavoidingly introduces singularities~\cite{Penrose:1964wq,Senovilla:2014gza}, and it struggles to fit in any quantum description~\cite{Shomer:2007vq}. 
From the observational perspective, the theory also struggles to explain certain phenomena without additional assumptions, such as the late-time acceleration of the universe (without an unnaturally small cosmological constant)~\cite{SupernovaSearchTeam:1998fmf,SupernovaCosmologyProject:1998vns}, the rotation curves of the galaxies (without introducing dark matter)~\cite{Sofue:2000jx,Bertone:2016nfn}, and the matter-antimatter asymmetry of the universe (without satisfying the Sakharov conditions)~\cite{Canetti:2012zc}. 
In response to the issues above, significant amount of theoretical work has gone into modifications to GR, and there is a growing need for testing GR against these modified gravity theories through lab experiments and astrophysical observations. 

The successful observation of gravitational waves (GWs) from compact binary coalescences (CBCs) in 2015 started a new era for testing GR~\cite{LIGOScientific:2016aoc,LIGOScientific:2016lio}. Unlike previous observational tests carried out in the Solar System~\cite{Will:2014kxa} or with binary pulsars~\cite{Stairs:2003eg}, these GW signals probe the gravitational interaction in a regime that involves both dynamical and strong gravitational fields~\cite{Yunes:2016jcc,Carson:2020rea}. 
Currently, there are $\sim100$ confirmed CBC events in the open GW transient catalogs published by the LIGO-Virgo-KAGRA (LVK) collaboration~\cite{LIGOScientific:2018mvr,LIGOScientific:2020ibl,LIGOScientific:2021usb,KAGRA:2021vkt} and by other researchers~\cite{Nitz:2018imz,Magee:2019vmb,Zackay:2019tzo,Nitz:2020oeq,Venumadhav:2019lyq,Zackay:2019btq,Nitz:2021uxj,Nitz:2021zwj}. More events are expected as the LVK collaboration is now operating a fourth observing run (O4) and a fifth run (O5) is scheduled to begin in 2027 with increased sensitivity~\cite{KAGRA:2013rdx}. 
These future observations may also result in the detection of fainter signals with higher signal-to-noise ratios (SNRs), as the instruments continue to get upgraded. 

Tests of GR with GW data generally fall into two categories: consistency tests and parametrized tests~\cite{LIGOScientific:2021sio}. 
Consistency tests, such as the residual test and the inspiral-merger-ringdown consistency test, search for possible violations of GR by asking how consistent the observed signal is with the prediction of GR. 
Parametrized tests, on the other hand, search for possible non-GR effects that are encoded by certain parameters in the gravitational waveform model. 
Although these parameters are related to specific non-GR theories, parametrized tests can still be theory-agnostic, as long as the parametrization covers the predictions of many non-GR models. 
One successful approach is provided by the parametrized post-Einstein (ppE) framework~\cite{Yunes:2009ke,Cornish:2011ys,Chatziioannou:2012rf,Sampson:2013lpa}, which introduces small deformations to the GR waveform that are motivated by generic, theoretical considerations and by a broad class of non-GR theories. 

The ppE framework is the general framework that the LVK's ``parametrized inspiral tests of GR'' is built from~\cite{Arun:2006yw,Mishra:2010tp,Li:2011cg,Agathos:2013upa,LIGOScientific:2016lio,Meidam:2017dgf,LIGOScientific:2018dkp,LIGOScientific:2019fpa,LIGOScientific:2020tif,Abbott:2020jks,LIGOScientific:2021sio,Mehta:2022pcn}. In these tests, the deformation is given by the leading-order modification to the post-Newtonian (PN) expansion\footnote{
This is an expansion of the inspiral waveform in powers of the binary velocity, which can be reexpressed in terms of the GW frequency through the PN version of Kepler's third law~\cite{Blanchet:1995ez}. 
} of the phase of the GR waveform. 
Such a prescription introduces two parameters: a ppE exponent or \textit{ppE index} representing the PN order at which the modification first takes place, and a \textit{ppE coefficient} representing the magnitude of the deformation at that given PN order. 
In terms of a physical interpretation, the ppE index maps to the type of non-GR modification one wishes to consider, and the ppE coefficient maps to the strength of the modification, including the coupling constants of modified gravity theories~\cite{Yunes:2009ke,Yunes:2016jcc,Tahura:2018zuq,Perkins:2020tra}. 
To run ppE or parametrized inspiral tests on GW data, one fixes the ppE index at a chosen value and estimates the ppE coefficient via Bayesian inference. The test is then made theory-agnostic by repeating the estimation for all ppE indices consistent with a non-GR theoretical prediction~\cite{LIGOScientific:2016lio,LIGOScientific:2018dkp,LIGOScientific:2019fpa,LIGOScientific:2020tif}.

Despite being standard, the ppE or parametrized inspiral test is unsatisfying for two reasons. 
First, the ppE test is not truly theory-agnostic. 
The ppE deformation applied to an inspiral waveform only captures modification at leading PN order. This means the modification must admit a PN expansion in the first place. Counterexamples exist, like the GWs generated by a binary in massive scalar-tensor theories~\cite{Alsing:2011er,Berti:2012bp,Liu:2020moh} or with dark-photon interactions~\cite{Alexander:2018qzg}, in which non-GR modifications activate suddenly, and thus, cannot be represented as a simple power of velocity. 
Even for modifications that are PN-expandable, a modeling bias can exist when certain higher-PN-order effects are omitted. 
Although one may formally add higher PN order terms to the ppE deformation, they must be accompanied with more coefficient parameters. 
Bayesian parameter estimation with this expanded parameter space will lead to uninformative results unless a certain marginalization procedure is implemented~\cite{Perkins:2022fhr}.
The second reason ppE inspiral tests are not ideal is their high computational expense, because of the need for repeating parameter estimation for many ppE indices. This issue may be mitigated by accelerating each estimation run with a hybridized sampling technique~\cite{Wolfe:2022nkv}, but still, one must repeat the analysis for many ppE indices.
A transdimensional reversible-jump Markov-chain Monte-Carlo algorithm can alleviate the latter~\cite{Green:1995mxx,Sampson:2013lpa}, but at an increased computational cost. 
We also note that a principal component analysis allows multiple PN terms to be estimated at once to search for potential deviations from GR~\cite{Saleem:2021nsb,Shoom:2021mdj,Pai:2012mv}, but the deviations constrained or found this way cannot be mapped back to any specific non-GR theory, and hence, imply nothing about fundamental physics.

In this paper, we attempt to address both of these weaknesses by extending the ppE parametrization for inspiral tests through a new \textit{neural post-Einsteinian} (npE) framework.
The framework is built by exploiting variational autoencoders (VAE)~\cite{1312.6114}, a type of neural network that has recently found applications in various GW studies, including glitched identification~\cite{Sankarapandian:2021qun}, waveform modeling~\cite{Liao:2021vec,Whittaker:2022pkd}, and parameter estimation~\cite{Gabbard:2019rde}. 
We use a VAE to find a latent space to parametrize GR deformations, using the ppE inspiral modifications as training data. After training, the VAE can use the learned latent space to generate waveform modifications that continuously interpolate between ppE waveforms with different ppE indices in a physics informed way. 
This means that testing GR with the npE parametrization will naturally explore a larger set of non-GR effects.
The unification of various non-GR effects into a single continuous parameter space also allows for computationally-efficient parameter estimation. 

We study the performance of the npE framework to test GR with simulated GW signals, using the anticipated LVK O5 sensitivity. 
First, we verify that the npE framework can be used to search for a variety of non-GR effects with a single parameter estimation run. For the injected signals, we consider not only modifications following the exact ppE model, but also those going beyond the ppE description, including higher-PN-order corrections in Einstein-dilaton-Gauss-Bonnet (EdGB) theory~\cite{Shiralilou:2021mfl}, and non-PN-expandable effects due to dark-photon interactions~\cite{Alexander:2018qzg}. We show that the npE framework is able to detect both types of non-GR deviations in all cases. This validation cements the npE framework as a new and promising tool to test GR with current and future GW observations. 

The rest of this paper is organized as follows. 
In Sec.~\ref{sec:tgr}, we review the theoretical and technical background for parametrized tests of GR with GW data. We summarize the ppE inspiral tests, and discuss their weaknesses. 
In Sec.~\ref{sec:npe}, we describe how we use the VAE to obtain the npE parametrization. In particular, we briefly review the formulation of VAEs in general in Sec.~\ref{sec:vae}, sketch out the general design of the npE parametrization in Sec.~\ref{sec:npe_sketch}, and specify the implementation of each component of the npE framework in the subsections that follow. 
In Sec.~\ref{sec:test_npe}, we showcase the npE test with simulated GW signals. We start by describing the GW sources that we study and the Bayesian inference procedure in Sec.~\ref{sec:test_setup}. 
The remaining subsections describe various injections and the npE recoveries. 
In Sec.~\ref{sec:constrain_gr}, we inject signals in GR and investigate the use of the npE framework to constrain non-GR deviations. In Sec.~\ref{sec:detect_ppe}, we inject signals with ppE modifications, and compare the npE and ppE performance when searching for these modifications. 
In Sec.~\ref{sec:detect_higher_pn}, we inject a signal in EdGB theory with modifications beyond leading PN order, and show that the npE parametrization recovers such a signal better than ppE templates.
In Sec.~\ref{sec:detect_hidden_sector}, we inject signals with modifications from the non-PN-expandable dark-photon interactions in the hidden sector, and show that the npE framework can handle such a case reasonably well.
Finally, in Sec.~\ref{sec:conclusions}, we draw conclusions and point to future work. We note that, this paper is prepared with extensive background information for an audience composed of members from various communities. Readers experienced in GW modeling and Bayesian parameter estimation may wish to skip Sec.~\ref{sec:cbc}, those experienced in the ppE framework may wish to skip Sec.~\ref{sec:ppe}, and those experienced in deep learning with the VAE may wish to skip Sec.~\ref{sec:vae}.
The main ideas behind the npE framework and our particular contributions begin in Sec.~\ref{sec:npe_sketch}.

Henceforth, we use geometric units, in which $G=1=c$. 
We reserve $|\cdot|$ for taking the absolute value of a number or a scalar, and use $\lVert\cdot\rVert$ for the $L_2$ norm of a given quantity.
Vectors will either be indexed or denoted by $\vec{(\cdot)}$ or $\hat{(\cdot)}$. 
In particular, we use Greek letters $(\alpha,\beta,\gamma)$ to index a vector of physical waveform parameters, Latin letters $(i,j,k)$ to index a vector in the npE parameter space, and $(\ell,m,n)$ to index the coefficients in a PN expansion (or other expansions alike). 
We reserve Latin letters $(a,b,c)$ for indexing the elements of the training set of the neural networks. We also reserve the symbol $\vec{(\cdot)}$ for data either from or comparable to an element of the training set, assuming the data has multiple components. The notation $\hat{(\cdot)}$ further refers to the $L_2$ normalized version of a $\vec{(\cdot)}$ vector. 
These conventions will be reiterated throughout the paper when they are first introduced. 

\section{Parametrized Theory-agnostic Tests of GR with GW Data} \label{sec:tgr}
In this section, we review key concepts and tools in GW science that are needed before introducing the npE framework. 
We review the modeling of CBC signals in GR, as well as the data analysis procedure used to extract information from these signals. 
Then, we will describe the ppE framework for inspiral tests of GR, and comment on its weaknesses to motivate the npE framework.

\subsection{CBC process and GW parameter estimation} \label{sec:cbc}
CBCs are the only sources of GW signals observed by current detectors to date. 
A CBC event can be divided into three stages: (i) the inspiral, (ii) the merger, and (iii) the ringdown. 
The first stage begins with two compact objects orbiting each other. Due to GW radiation-reaction, the distance between the objects gradually decreases. This process accelerates over time and eventually the objects enter the second stage, where they plunge and merge. The third stage then follows, as the remnant rings down and stabilizes. 
Detailed modeling of the CBC waveform involves solving the Einstein equation in the dynamical and strong-field regime, in which one would like to employ numerical relativity (NR). However, for the inspiral stage, where the velocity is relatively small and the field is relatively weak, an analytical derivation of the waveform is possible with a PN expansion~\cite{Blanchet:1995ez}.

Discussion in later sections will focus on quasicircular and nonprecessing, binary black holes (BBH), and will use the IMRPhenomD waveform~\cite{Husa:2015iqa,Khan:2015jqa} to describe GWs from such sources in GR. The IMRPhenomD waveform models the GW response function in the frequency domain via
\begin{align}
    \tilde{h}(f)=A(f)\,e^{i\Phi(f)},
\end{align}
where $A(f)$ and $\Phi(f)$ are the frequency-domain amplitude and phase, respectively, given as piecewise functionals, some pieces of which are fitted to NR data. The amplitude $A(f)$ and the phase $\Phi(f)$ each contains three pieces, associated with each of the three stages of CBC described above. At low frequencies, the functionals describe the inspiral, at intermediate frequencies they describe an intermediate stage prior to merger, and at high frequencies they describe the merger and ringdown.
In particular, for the inspiral piece, $A(f)$ and $\Phi(f)$ take the form of a PN expansion, with coefficients analytically calculated up to $3$PN and $3.5$PN, respectively, and higher PN order terms fitted to NR data.

For concreteness, let us write down the IMRPhenomD inspiral phase for later reference:
\begin{align}
    \Phi(f) =& \,2\pi ft_c - \phi_c - \frac{\pi}{4} \notag\\
    &+ \frac{3}{128\eta}(\pi Mf)^{-5/3} \sum_{n=0}^{11} \phi_n(\Xi_\alpha)\,(\pi Mf)^{n/3}, \label{eqn:cbc_imr_insp_phase}
\end{align}
where $t_c$ and $\phi_c$ are the time and phase of coalescence, respectively. 
The total mass $M$ and the symmetric mass ratio $\eta$ are defined as
\begin{align}
    M=m_1+m_2,\quad \eta=\frac{m_1m_2}{(m_1+m_2)^2} = \frac{q}{(1 + q)^2},
\end{align}
where $m_{1,2}$ are the component BH masses and $q = m_2/m_1 < 1$ is the mass ratio.
The PN expansion is carried out in powers of the product of the total mass and frequency, where a term proportional to $(\pi Mf)^{n/3}$ is said to be of $(n/2)$PN order. Each PN coefficient $\phi_n$ is calculated or fitted as a function of the intrinsic BBH parameters
\begin{align}
    \Xi_\alpha=(M,\,\eta,\,\chi_1,\chi_2),
\end{align}
where $\chi_{1,2}$ are the dimensionless component spins projected onto the direction of the orbital angular momentum. More details about the $\phi_n$ coefficients can be found in Appendix A and C in~\cite{Khan:2015jqa}. 
As a final remark, the frequency range for the use of Eq.~\eqref{eqn:cbc_imr_insp_phase} is 
\begin{align}
    Mf<0.018. \label{eqn:cbc_imr_insp_phase_cutoff}
\end{align}
Beyond this range are the intermediate and the merger-ringdown regions, for which different empirical ansatz are fitted. However, at $Mf=0.018$ (as well as the counterpart separating the intermediate and the merger-ringdown regions), the different pieces of the phase function are stitched with $C^1$ smoothness.

With the waveform model at hand, one may analyze the GW data collected by detectors and recover the properties of the binary merger. Bayesian inference is applied to estimate the parameters of the binary, such as the masses, the spins, the distance, the sky location, etc, in terms of their posterior distribution. 
According to Bayes' theorem, 
\begin{align}
    p(\{\Theta_\alpha\}|d)\propto p(d|\{\Theta_\alpha\})\,p(\{\Theta_\alpha\}), \label{eqn:cbc_posterior}
\end{align}
where $p(\{\Theta_\alpha\}|d)$ is the posterior for the set of CBC parameters $\Theta_\alpha$ given the GW data $d$, $p(\{\Theta_\alpha\})$ is the prior of the parameters, and $p(d|\{\Theta_\alpha\})$ is the likelihood function. 
Assuming that the detector noise is additive and Gaussian, the likelihood can be further expressed as:
\begin{align}
    p(d|\{\Theta_\alpha\}) \propto e^{-\frac{1}{2} \langle d-h(\Theta_\alpha)\, |\, d-h(\Theta_\alpha) \rangle}, \label{eqn:cbc_likelihood}
\end{align}
where $h$ is the detector response given by the waveform model, and $\langle\cdot|\cdot\rangle$ is an inner product defined as
\begin{align}
    \langle A|B \rangle = 4\mathrm{Re} \int_{f_\mathrm{low}}^{f_\mathrm{high}} \frac{\tilde{A}(f)\,\tilde{B}(f)^*}{S_n(f)}\,df,
\end{align}
where $\tilde{(\cdot)}$ denotes Fourier transformation, $(\cdot)^*$ denotes complex conjugation, $S_n(f)$ is the one-sided power spectral density of the detector noise, and $f_\mathrm{low}$ ($f_\mathrm{high}$) is the lowest (highest) frequency considered for the GW data. 
If the observation is made by a network of detectors, Eq.~\eqref{eqn:cbc_likelihood} has to be evaluated for each detector and results have to be summed up to obtain the total log likelihood function. 

The numerical estimation of the posterior can be carried out with sampling algorithms, such as Markov-Chain Monte Carlo and nested sampling methods. Due to the high dimensionality of the GW parameter space (e.g.~11 free parameters in IMRPhenomD), parameter estimation can be computationally expensive, typically requiring the likelihood to be evaluated tens of millions of times before the algorithm converges~\cite{Wolfe:2022nkv}. Methods to accelerate the computational process of posterior sampling remain an active area of research in GW data science (see e.g.~\cite{Canizares:2014fya,Smith:2016qas,Zackay:2018qdy,Morisaki:2021ngj,Williams:2021qyt,Dax:2021tsq,Dax:2022pxd,Wong:2023lgb,Veitch:2014wba,Ashton:2021anp,Cornish:2021wxy,Islam:2022afg,Roulet:2022kot,Lee:2022jpn,Lange:2018pyp,Wofford:2022ykb}).

\subsection{GW tests of GR with the ppE Framework} \label{sec:ppe}
The previous subsection has briefly summarized the study of astrophysical properties of binary mergers with GW data. The same GW parameter estimation scheme also applies to parametrized tests of GR, which are typically carried out by introducing a modified waveform that admits the GR one as a nested model. 

Consider, for example, a specific modified-gravity theory that recovers GR when its coupling constant goes to zero. In particular, let us assume that the waveform in this theory can be derived analytically or numerically, and that the waveform is characterized by the BBH astrophysical parameters and the coupling constant of the theory. Bayesian parameter estimation can then be carried out as usual, with only the simple modification of adding the coupling constant to the parameter array $\Theta_\alpha$. 
Given a detected CBC event, one may then run GW parameter estimation with the modified waveform and the augmented parameter set, and a deviation from GR can be claimed if the marginalized posterior of the coupling constant is statistically incompatible with zero. 
On the other hand, if GR is not rejected by the posterior, a constraint on the modified-gravity theory can be obtained with the upper bound of the coupling constant at a certain credible level.

We note that the test described above is theory-specific. A more theory-agnostic test can be built from the ppE framework, like those performed by the LVK~\cite{LIGOScientific:2016lio,LIGOScientific:2018dkp,LIGOScientific:2019fpa,LIGOScientific:2020tif,Abbott:2020jks,LIGOScientific:2021sio}.
In this scheme, the GR waveform is deformed by terms that are motivated by generic theoretical considerations and cover a broad class of non-GR theories. 
Let us then follow the prescription of~\cite{Yunes:2016jcc} and consider only modifications to the frequency-domain phase of the inspiral GW. The ppE waveform can then be written as
\begin{align}
    \tilde{h}_\mathrm{ppE}(f) =&\, \tilde{h}_\mathrm{GR}(f)\,e^{i \Phi_\mathrm{ppE}(f)}, \\
    \Phi_\mathrm{ppE}(f) =&\, \beta_\mathrm{ppE}\, (\pi\mathcal{M}f)^{b_\mathrm{ppE}/3}, \label{eqn:ppe_phi}
\end{align}
where $\tilde{h}_\mathrm{GR}(f)$ is the GR waveform (e.g.~IMRPhenomD), $b_\mathrm{ppE}$ is the ppE index parameter, $\beta_\mathrm{ppE}$ is the ppE coefficient parameter, and $\mathcal{M}=M\eta^{3/5}$ is the chirp mass of the binary. 

The physical motivation of the ppE framework requires the index $b_\mathrm{ppE}$ to take only integer values, with each integer featuring certain non-GR effects. This formally aligns $\Phi_\mathrm{ppE}$ with the $\left(\frac{b+5}{2}\right)$PN term in the GR phase [see e.g.~Eq.~\eqref{eqn:cbc_imr_insp_phase}]. 
Therefore, Eq.~\eqref{eqn:ppe_phi} models deviations from GR that capture the leading-order PN correction in the GW phase. 
In Table~\ref{tab:ppe_indices_theories}, we provide a list of ppE indices with the effects and example theories they may represent. 
Mappings between $\beta_\mathrm{ppE}$ and the physical constants (the coupling strength, mass, etc) in each theory can be further found in~\cite{Yunes:2016jcc,Carson:2019yxq}.
\begin{table*}[htbp]
    \centering
    \begin{tabular}{c|c|c|c}
        \hline
        \hline
        $b_\mathrm{ppE}$ & Order & Theoretical Effect & Example Theories \\ 
        \hline
        \multirow{2}{*}{$-13$} & \multirow{2}{*}{$-4$PN} & \multirow{2}{*}{Anomalous Acceleration} & RS-II Braneworld~\cite{Randall:1999ee,Randall:1999vf} \\
        & & & Varying-G~\cite{1937Natur.139..323D,Yunes:2009bv} \\
        \hline
        \multirow{2}{*}{$-7$} & \multirow{2}{*}{$-1$PN} & \multirow{2}{*}{Scalar Dipolar Radiation} & Einstein-dilaton Gauss-Bonnet~\cite{Metsaev:1987zx,Maeda:2009uy,Yunes:2011we,Yagi:2011xp}  \\
        & & & Scalar-Tensor Theories~\cite{Jacobson:1999vr,Horbatsch:2011ye} \\
        \hline
        \multirow{2}{*}{$-5$} & \multirow{2}{*}{$0$PN} & \multirow{2}{*}{Modified Quadrupolar Radiation} & Einstein-\AE ther~\cite{Jacobson:2000xp,Jacobson:2008aj}  \\
        & & & Khronometric~\cite{Blas:2009qj,Blas:2010hb} \\
        \hline
        $-3$ & $1$PN & Second Order Dispersion & Massive Gravity~\cite{Will:1997bb,Rubakov:2008nh,Hinterbichler:2011tt,deRham:2014zqa} \\
        \hline
        \multirow{3}{*}{$-1$} & \multirow{3}{*}{$2$PN} & Scalar Quadrupolar Radiation & \multirow{3}{*}{Dynamical Chern-Simons~\cite{Alexander:2009tp,Yagi:2011xp}} \\
        & & Scalar Dipole Force & \\
        & & Quadrupole Moment Deformation & \\
        \hline
        \hline
    \end{tabular}
    \caption{Mapping between the ppE index $b_\mathrm{ppE}$ and example non-GR effects/theories. Adapted from~\cite{Yunes:2016jcc}.}
    \label{tab:ppe_indices_theories}
\end{table*}

To run a ppE or parametrized test of GR, one chooses a list of integer $b_\mathrm{ppE}$'s and estimate $\beta_\mathrm{ppE}$ for each one of them. 
If any of the $\beta_\mathrm{ppE}$ posteriors is statistically incompatible with zero, a deviation from GR can be claimed under the physical mechanisms implied by the corresponding $b_\mathrm{ppE}$. If GR is not rejected by the posterior, the constraint on each $\beta_\mathrm{ppE}$ can be mapped to constraints on the theories associated. 
The ppE framework is relatively efficient, because a large number of non-GR theories are ``compressed'' into a smaller (and finite) set of ppE indices. Moreover, if new theories are developed in the future, chances are that they can be represented in the ppE way at leading PN order, and that constraints on those ppE indices may already have been obtained.
% We also note that, the ppE framework is always based on a GR waveform $\tilde{h}_\mathrm{GR}(f)$, which may contain modeling bias for its own objective. In that case, the ppE framework may be used to study the waveform systematics, or to discover exotic matter effects that are not considered for standard BHs and NSs in the vacuum. 

Despite having been widely adopted,~\cite{Arun:2006yw,Mishra:2010tp,Li:2011cg,Agathos:2013upa,LIGOScientific:2016lio,Meidam:2017dgf,LIGOScientific:2018dkp,LIGOScientific:2019fpa,LIGOScientific:2020tif,Abbott:2020jks,LIGOScientific:2021sio,Mehta:2022pcn} the ppE framework can be unsatisfactory for two main reasons. 
First, ppE tests are not truly theory-agnostic. Equation~\eqref{eqn:ppe_phi} only makes sense for theories that admit a PN expansion for their description of the inspiral. Counterexamples exist, like the massive scalar-tensor theories~\cite{Alsing:2011er,Berti:2012bp,Liu:2020moh} and theories with dark-photon interactions in the hidden sectors of matter~\cite{Alexander:2018qzg}, in which this is not the case. 
In these theories, the gravitational dynamics is affected by the competition between GW quadrupole emission, occurring at the binary scale, and scalar emission, which can be dynamically activated at a certain (Compton wavelength) scale associated with the mass of the scalar field. A typical consequence of this is that the phase of the inspiral waveform acquires modifications that are proportional to a Heaviside step function in frequency (to be discussed in more detail in Sec.~\ref{sec:detect_hidden_sector}). This step function introduces sudden modifications to the waveform phase, which cannot be described as a Taylor series in velocity. Moreover, even for theories that admit PN-expandable modifications, a modeling bias can exist when certain higher-PN-order effects are omitted.
Although one may formally add higher PN order terms to Eq.~\eqref{eqn:ppe_phi}, they must be accompanied with more coefficient parameters, and the prior range of each coefficient must be carefully regularized in order for Bayesian parameter estimation to produce informative results after marginalizing over the higher-PN-order terms~\cite{Perkins:2022fhr}.

The second reason ppE inspiral tests are not ideal is their high computational expense. Although the ppE framework is more efficient than separately testing each specific theory available in the literature, one still has to run Bayesian parameter estimation for every element in a list of ppE indices, repeating parameter estimation many times. 
This issue may be mitigated by a sampling technique that hybridizes one expensive estimation in GR with a set of cheaper estimations that include ppE deviations~\cite{Wolfe:2022nkv}, but still one must repeat the analysis for many ppE indices. Though the latter can be alleviated with a transdimensional reversible-jump Markov-chain Monte Carlo algorithm that enables sampling across parameter spaces with varying dimensions~\cite{Green:1995mxx,Sampson:2013lpa} (such as transitions between GR models and non-GR ppE models), such an algorithm is complex and may elevate computational expenses.

Previous to this work, a principal component analysis has been shown to be effective at extracting information from parameter estimation studies with multiple non-GR PN terms included~\cite{Saleem:2021nsb,Shoom:2021mdj,Pai:2012mv}. This approach seems to address both weaknesses of the ppE framework (if non-PN-expandible theories are ignored). 
However, the information extracted by such a principal component analysis is always encoded in linear combinations of all PN terms in the estimate. Such a combination may be used to identify a deviation from GR (or the presence of an astrophysical environment), but it cannot be mapped back to any specific non-GR theories.

\section{NpE Framework} \label{sec:npe}
In this section, we describe the npE framework that extends the ppE parametrization and addresses the weaknesses of the ppE tests. 
This is achieved using neural networks that exploit the VAE architecture. Neural networks are universal function approximators~\cite{Cybenko:1989iql}. A VAE adopts neural networks in its substructure to find an effective parametrization of the given data. 
In our case, this allows us to build the npE waveform template to search for deviations from GR that are not required to follow a human-designed parametrization (e.g.~$b_\mathrm{ppE}$ and $\beta_\mathrm{ppE}$).
We will begin with a brief introduction to VAEs, as well as a sketch of how VAEs will be utilized as the base of the npE parametrization. 
We will then describe our implementation of the npE framework, and give a preliminary analysis on the training results of these neural networks. 

\subsection{VAE} \label{sec:vae}
Neural networks are numerical approximants to complicated functions~\cite{Cybenko:1989iql}. 
The defining structure of a neural network is a sequence of alternating linear and nonlinear mappings. 
The nonlinear mapping at each level of the network, also known as a ``hidden layer,'' is implemented as a nonlinear activation (e.g.~a sigmoid function) that acts element-wise on each component of the input. Such an activation is analogous to a collection of neurons reacting to some neurological signals in synchrony. 
The linear mappings between adjacent layers are then analogous to the connections between neurons, and they contain parameters to be fitted by minimizing a loss function that evaluates the difference between the network output and the expected outcome.
This process of minimization is also known as ``learning'' or ``training,'' and it is usually carried out through methods based on stochastic gradient descent. 
For more general information about neural networks, the readers may refer to~\cite{Goodfellow-et-al-2016}. 

A VAE~\cite{1312.6114} is a class of neural network designed to (i) find a representation for the underlying pattern of a given dataset, and (ii) generate samples from that representation that mimic these data. More precisely, a VAE \textit{encodes} a training dataset into a dimensionally-reduced ``latent space'' by \textit{automatically} minimizing a certain loss function. The \textit{variational} part of the VAE enters through the choice of loss function, which guarantees continuity across the latent space, i.e.~two infinitesimally-separated points in the latent space lead to decoded data that is also similarly infinitesimally separated.

\begin{figure}
    \centering
    \includegraphics[width=0.48\textwidth]{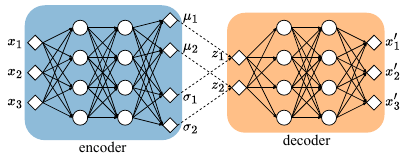}
    \caption{Diagram of a VAE, composed of an encoder subnetwork (blue-shaded) mapping from the data $\vec{x}$ to the latent representation $(\mu_i,\sigma_i)$, and a decoder subnetwork (orange-shaded) mapping from a latent point $z_i$ to the reconstructed data $\vec{x}'$. The diamonds are the input or output of a network. The circles represent ``neurons'' of nonlinear functions in the hidden layer. The solid arrows represent linear mappings, with free parameters to be fitted by minimizing the VAE loss function. The dashed arrows represent the random sampling of $z_i\sim\mathcal{N}(\mu_i,\sigma_i)$.
    This diagram showcases an example, in which the data is 3-dimensional, the latent space is 2-dimensional, and the encoder and decoder are each fully connected with 2 hidden layers of width 4.
    }
    \label{fig:vae}
\end{figure}

Let us then dig deeper into the construction of a VAE. 
This kind of neural network is composed of two subnetworks: an encoder network and a decoder network (see Fig.~\ref{fig:vae}).
The encoder takes some data $\vec{x}=(x_1,x_2,x_3,\ldots)$ and finds a distributional representation of this data in a latent space. More precisely, in the standard VAE formalism, the encoder outputs a normal distribution with mean $\mu_i$ and standard deviation $\sigma_i$ for each dimension $i$ of the latent space. For example, if the latent space is 2-dimensional, then there are two normal distributions with mean and standard deviations $(\mu_1,\sigma_1)$ and $(\mu_2,\sigma_2)$.
The decoder, then, takes a sample $z_i$ in the latent space and generates some $\vec{x}'=(x_1',x_2',x_3',\ldots)$ in such a way as to resemble $\vec{x}$, which is achieved by minimizing a loss function. 

The loss function for training of a VAE is
\begin{align}
    L_\mathrm{VAE}=\sum_a \left[ L_\mathrm{recon}\big(\vec{x}^{(a)},\vec{x}'(z_i^{(a)})\big) + \kappa\,D_\mathrm{KL}\big(\mu_i^{(a)},\sigma_i^{(a)}\big) \right], \label{eqn:vae_loss_template}
\end{align}
where $\vec{x}^{(a)}$ is the $(a)$th element in the training set. For each $\vec{x}^{(a)}$, we draw a sample $z_i^{(a)}$ from the latent space created by the encoder, with mean $\mu_i^{(a)}$ and standard deviation $\sigma_i^{(a)}$. This sample $z_i^{(a)}$, then, allows the decoder to generate an $\vec{x}^{\prime(a)}$. For a decoded $\vec{x}^{\prime(a)}$ to resemble the element $\vec{x}^{(a)}$ of the training set, we must minimize $L_\mathrm{VAE}$, because $L_\mathrm{recon}$ penalizes the difference between $\vec{x}^{\prime(a)}$ and $\vec{x}^{(a)}$ (with a choice of an appropriate metric), while $D_\mathrm{KL}$ penalizes the difference between the encoded distribution of the latent space and a prior distribution. 
More specifically, $D_\mathrm{KL}$ is the Kullback-Leibler (KL) divergence between $\mathcal{N}(\mu_i^{(a)},\sigma_i^{(a)})$ and a prior distribution of the latent representation, which one usually chooses to be $\mathcal{N}(0,1)$ by default. 
This KL term regularizes the VAE model, and its importance in the total loss function is tuned by the coefficient $\kappa$.

After training, the decoder becomes a continuous model for the overall structure and variation of the original data. This model takes the latent parameters $z_i$ as input, and it outputs an approximation to the original data. Therefore, after training, the encoder is not needed any longer for our purposes, but it could be used as an approximation for the inverse of the decoder. The model is continuous on the latent parameters in the following sense. When the model is evaluated at the mean $\mu_i^{(a)}$, it generates an output $\vec{x}^{\prime}(\mu_i^{(a)})$ that is close to $\vec{x}^{(a)}$. If we now evaluate the model at a point in the latent parameter space that is shifted from the mean by some small amount $\Delta_i$, the output generated $\vec{x}^{\prime}(\mu_i^{(a)}+\Delta_i)$ will remain close to $\vec{x}^{\prime}(\mu_i^{(a)})$, and thus, close to $\vec{x}^{(a)}$. This is ensured by the fact that $L_\mathrm{recon}$ remains small only if $\vec{x}^{\prime}(\mu_i^{(a)}+\Delta_i)$ is close to $\vec{x}^{(a)}$ for any $\Delta_i$ smaller than $\sigma_i$. The inclusion of $D_\mathrm{KL}$ in the loss function regularizes the latent representation by pushing the encoder output to match the prior distribution. 

\begin{center}
    \textbf{Reason for using a VAE}
\end{center}

The above features make the VAE a useful tool for theory-agnostic tests of GR with GW data, which may overcome some of the weaknesses of the ppE framework. 
A neural-enhanced ppE framework would then arise from the training of a VAE on GW models that contain as many GR modifications as can be modeled. After training, the decoder would be used for parameter estimation, with non-GR parameters taken directly from the latent space. This way, one single Bayesian parameter estimation study across the latent space would be sufficient to test GW data against \textit{all} theories of gravity that have been modeled. 
Furthermore, because the latent space is continuous, the gap between the latent training data points (i.e.,~gaps in \textit{theory space}) can still be used to construct a VAE model, thus automatically extending the non-GR waveforms beyond known modified theories. Tests of GR with this npE model would therefore be naturally more theory-agnostic than those limited by the ppE scheme. 

\subsection{A sketch of the npE framework} \label{sec:npe_sketch}
Now that we have provided an introduction to neural networks and VAEs, let us discuss how we will implement these machine-learning tools to develop an npE model.
Similar to the ppE framework, let us consider a generic phase modification to the GR \textit{inspiral} signal in the frequency domain:
\begin{align}
    \tilde{h}_\mathrm{mod}(f) =&\, \tilde{h}_\mathrm{GR}(f)\,e^{i \Phi_\mathrm{mod}(f)}\,; \label{eqn:npe_generic_mod}
\end{align}
we only consider here inspiral modifications to GR, leaving the modeling of non-GR modifications to the intermediate and merger-ringdown stages of the IMRPhenom models to future work. 
In the ppE framework, $\Phi_\mathrm{mod}(f)=\Phi_\mathrm{ppE}(f;\mathcal{M},b,\beta)$. 
In our npE framework, we start with the following generic form of the non-GR phase model:
\begin{align}
    \Phi_\mathrm{mod}(f)=\Phi_\mathrm{npE}(f;\Xi_\alpha,z_i),
\end{align}
where $z_i$ contains all parameters needed to describe the non-GR modifications. This modified GW phase in the frequency domain is then the quantity that we will develop a VAE to emulate. 

Because we intend to use the npE framework for tests of GR through Bayesian inference, we expect the vector $z_i$ in $\Phi_\mathrm{npE}$ to be accompanied by a simple prior that describes a uniform distribution of theories for modifications of different magnitude. 
An example of such a simple prior, and the choice we will make in this paper, is that of a flat distribution bounded by the unit hypersphere centered at $z_i=0$. We choose $z_i=0$ to represent GR, and $\lVert z_i\rVert=1$ to represent a ``prior boundary,'' associated with the largest modifications of interest. 
In this paper, we will require that the modifications to GR be small deformations (as in the ppE framework), and thus, we will not allow these corrections to be larger than the GR terms in the GW phase themselves; this EFT condition sets the size of the prior, which we shall refer to as the \textit{EFT prior boundary}\footnote{Note of caution is now necessary. By ``EFT boundary'', we do not here mean a cut-off boundary beyond which the energies probed are beyond the regime of validity of the effective theory. Rather, we mean a boundary at which the GR modifications cease to be perturbatively small on average over all source parameters.}.
With this in mind, the magnitude $\lVert z_i\rVert$ represents the size of the modification, $\Phi_\mathrm{npE}\propto\lVert z_i\rVert$, while the direction in the latent space $n_i=z_i/\lVert z_i\rVert$ represents the type of GR modification. 

The above considerations motivate the following choice for the npE phase model:  
\begin{align}
    \Phi_\mathrm{npE}(f;\Xi_\alpha,z_i)=\lVert z_i\rVert\,T(\Xi_\alpha,n_i)\,S(\bar{f};n_i). \label{eqn:npe_decomp}
\end{align}
We define the quantity $S$ as the \textit{shape function}, which will contain all the frequency dependence and encode the type of modification we wish to consider. 
We have here replaced $f$ with the dimensionless frequency $\bar{f}=Mf$, which scales better between different sources. 
We define the quantity $T>0$ as the \textit{scale function}, which ensures the prior bound at $\lVert z_i\rVert=1$ is the same for all sources. 
The actual functional forms of $S$ and $T$ are left undetermined for now, and they will be calculated later through the VAE algorithm. For now, however, let us consider the symmetry properties of $S$ and $T$. We wish for $\Phi_\mathrm{npE}$ to be odd under $z_i \to - z_i$, i.e.
\begin{align}
    \Phi_\mathrm{npE}(f;\Xi_\alpha,-z_i)=-\Phi_\mathrm{npE}(f;\Xi_\alpha,z_i)\,,\label{eqn:npe_parity}
\end{align}
This allows points along the $+ n_i$ and the $-n_i$ directions in the latent space to be interpreted as the same type of modification, assigning the same prior density in both directions.
One way to ensure this is the case is to require that $S$ and $T$ be an odd and even function of $n_i$, respectively. 

In our npE phase model, the shape function will be learned through a VAE, which we will train first. The scale function will be learned through a separate neural network, which we refer to as the secondary network and which we will train later, treating the latent parametrization as already completed by the trained VAE.
Such a separation of the training processes is beneficial when the prior boundary is chosen to reflect the most stringent constraints on theories of interest. If these constraints are tightened by future observations, one only needs to retrain the secondary network to update the scale function, while the VAE for the shape function does not necessarily require an update. 

\begin{figure}
    \centering
    \includegraphics[width=0.42\textwidth]{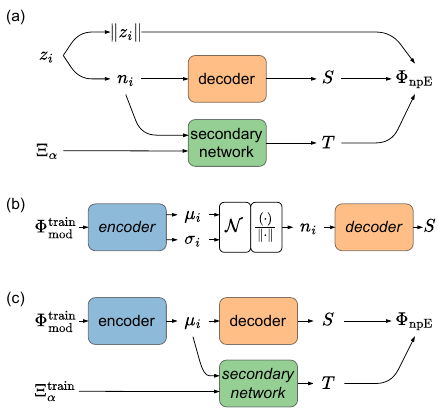}
    \caption{
    Ways to connect the npE networks to (a) model the npE phase modification $\Phi_\mathrm{npE}$ with the latent parameters $z_i$ and the source parameters $\Xi_\alpha$, (b) train the VAE for the shape function $S$ given a dataset of non-GR phase modifications $\Phi_\mathrm{mod}^\mathrm{train}$, and (c) train the secondary network for the scale function $T$ given $\Phi_\mathrm{mod}^\mathrm{train}$ and $\Xi_\alpha^\mathrm{train}$ in the same dataset.
    The quantity $n_i$ represents a unit vector in the latent space, and $\mu_i$ and $\sigma_i$ are the mean and standard deviation of the encoder output, respectively. 
    Each rounded rectangle represents an operation, either predefined (uncolored) or provided by a neural network (colored). For the predefined operations, $\mathcal{N}$ represents a random draw from a normal distribution, and $\frac{(\cdot)}{\lVert\cdot\rVert}$ means vector normalization. For the neural networks, the name of the network is italic if the network parameters are being fitted, and is roman if the network is treated as a known function and no parameter update takes place.}
    \label{fig:npe_connect}
\end{figure}

\begin{figure}
    \centering
    \includegraphics[width=0.49\textwidth]{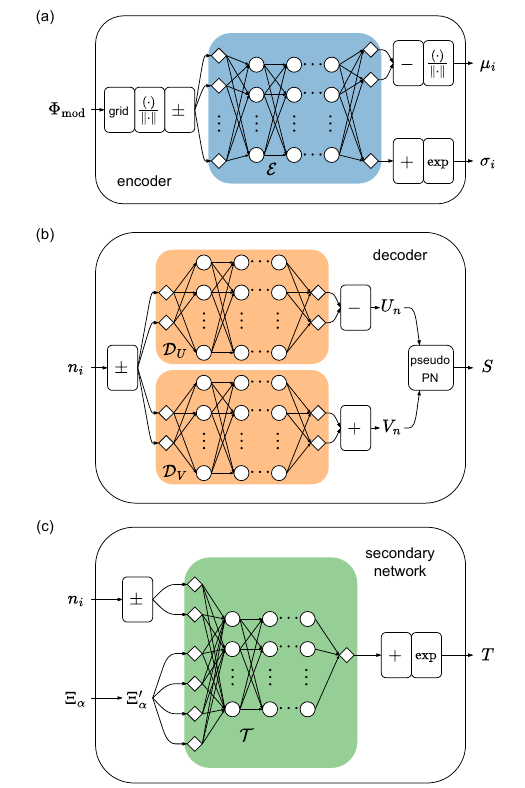}
    \caption{
    Internal structure of (a) the encoder, (b) the decoder, and (c) the secondary network, as one zooms into the colored rounded rectangles in Fig.~\ref{fig:npe_connect}. 
    These networks employ the fully-connected neural maps $(\mathcal{E},\mathcal{D}_U,\mathcal{D}_V,\mathcal{T})$ and wrap them with appropriate operations to condition their input and output.
    The neural maps are illustrated following the same format as in Fig.~\ref{fig:vae}. The actual dimension of these neural maps are significantly larger than what is shown here, which we indicate with ellipses. 
    The operators wrapping the neural maps are represented by uncolored rounded rectangles inside each network. For the notation of these operators, we have already introduced $\frac{(\cdot)}{\lVert\cdot\rVert}$ for vector normalization in Fig.~\ref{fig:npe_connect}. 
    In addition, ``$\exp$'' means the natural exponential, ``grid'' means discretizing a function of frequency in a frequency grid, and ``pseudo-PN'' refers to $\sum_n U_n \bar{f}^{V_n}$ to construct the shape function (see more details explained in Sec.~\ref{sec:shape_vae}). 
    The operations denoted by $\pm$, $+$ and $-$ are related to the symmetry considerations discussed in the text. In particular, $\pm$ creates two copies of the input, and add a minus sign to one of them. Once the two copies have both gone through a neural map, the $+$ ($-$) operation takes the sum (difference) of the two corresponding outputs, enforcing symmetrization (antisymmetrization) of the overall network. 
    Finally, in the secondary network, the inputting source parameters $\Xi_\alpha$ is coordinate-transformed into $\Xi'_\alpha=\big(\ln\mathcal{M},q,(\chi_1+\chi_2)/2,(\chi_1-\chi_2)/2\big)$ before being inserted into the neural map $\mathcal{T}$.
    }
    \label{fig:npe_networks}
\end{figure}

As a summary, Fig.~\ref{fig:npe_connect} shows how the encoder, the decoder and the secondary network are connected and how they can be used to carry out different tasks, while Fig.~\ref{fig:npe_networks} illustrates the internal structure of each network. Our discussion so far has covered Fig.~\ref{fig:npe_connect} (a), i.e.~how $\Phi_\mathrm{npE}$ is given by the decoder and the secondary network. The elements of other panels in these figures will be explained in the following subsections. In particular, we will specify, in order, the dataset for training, the detailed implementation of the VAE and the secondary network, the training procedure, and the training results. 

\subsection{Modified gravity dataset for the npE networks} \label{sec:data_set}
Our training and validation sets will be constructed from a modified gravity dataset, which will be created from a range of ppE models for non-GR frequency-domain GW phases generated by BBH sources.
Because ppE modifications with the same $b_\mathrm{ppE}$ do not differ in shape, we may create the dataset with only the largest modifications of interest,
\begin{align}
    \Phi_\mathrm{mod}^{(a)}(f)=\beta_\mathrm{ppE,max}^{(a)}\big(\pi\mathcal{M}^{(a)}f\big)^{b_\mathrm{ppE}^{(a)}/3}, 
\end{align}
where the superscript $(a)$ iterates over all elements in the dataset, i.e.~over all astrophysical parameters and all ppE theories of gravity considered. 

Here, we quantify the largest modification as
\begin{align}
    \beta_{\mathrm{ppE,max}}=\left\{\begin{aligned}
        &|\phi_{b+5}|\,\eta^{-(b+5)/5},&\,&b\geq-5, \\
        &|\phi_0|\,(\pi\mathcal{M}f_\mathrm{low})^{-(b+5)/3},&\,&b<-5,
    \end{aligned}\right. \label{eqn:npe_max_beta}
\end{align}
where $\phi_n$ is the $(n/2)$PN coefficient of the GR phase in Eq.~\eqref{eqn:cbc_imr_insp_phase}\footnote{The coefficient vanishes for $b=-4$ ($0.5$PN) and one may, for example, replace it with $\phi^\prime_{1}=\sqrt{|\phi_0\phi_2|}$.}, and $f_\mathrm{low}$ is the smallest frequency that we are sensitive to in the GW data.
For observations made during the O5 run, we choose $f_\mathrm{low}=10\,\mathrm{Hz}$. 
Equation~\eqref{eqn:npe_max_beta} implies that the ppE modifications we consider are small compared to GR effects in the observed data. 
An alternative to Eq.~\eqref{eqn:npe_max_beta} could be the maximal $\beta_\mathrm{ppE}$ associated with observational constraints on theories associated with $b_\mathrm{ppE}$. In this work, we choose Eq.~\eqref{eqn:npe_max_beta} as a proof of principle. 

The modified gravity dataset is then composed of elements that represent GW phases in different theories of gravity and produced by BBHs with different astrophysical properties. For the former, 
we choose odd ppE indices between $-13$ and $-1$, which cover all of the modified gravity predictions listed in Table~\ref{tab:ppe_indices_theories}.
For the BBH sources, we assume a population with masses $m_{1,2}$ distributed uniformly in the range $[5,\,30]\,M_\odot$. The spins are always aligned, and the population of $\chi_{1,2}$ is assumed to be uniform in the range $[-0.99,\,0.99]$. We randomly initiate 22,500 BBH sources for each ppE theory, creating a dataset of total size 157,500. 

To numerically represent $\Phi_\mathrm{mod}^{(a)}(f)$, we choose a grid of $f$ and evaluate $\Phi_\mathrm{mod}^{(a)}$ as a vector composed of phases at each grid point, i.e.,
\begin{align}
    \vec{\Phi}_\mathrm{mod}^{(a)}=\left(\Phi_\mathrm{mod}^{(a)}(f_1),\,\Phi_\mathrm{mod}^{(a)}(f_2),\,\Phi_\mathrm{mod}^{(a)}(f_3),\ldots\right), \label{eqn:npe_discretize}
\end{align}
In this work, we make a grid of $640$ points equally spaced in $\ln\bar{f}$, ranging from $\bar{f}_\mathrm{min}=0.0004$ to $\bar{f}_\mathrm{max}=0.018$. 
The choice of $\bar{f}_\mathrm{min}$ ensures that the low-frequency cutoff $f_\mathrm{low}=10\,\mathrm{Hz}$ is covered even for the smallest masses in the dataset. The $\bar{f}_\mathrm{max}$ above, on the other hand, serves as an inspiral cutoff for the IMRPhenomD model, taken from Eq.~\eqref{eqn:cbc_imr_insp_phase_cutoff}. 
The $640$-point grid, equally spaced in $\ln\bar{f}$, ensures that the ppE phase modifications in our modified gravity set are appropriately \textit{dense}, in the sense that 
\begin{align}
    \left| \Phi_\mathrm{mod}^{(a)}(f)-\Phi^{(a)}_\mathrm{mod,interp}(f) \right| \lesssim \frac{\pi}{10},\; \forall \; a \, {\textrm{and}} \,f,
\end{align}
where $\Phi^{(a)}_\mathrm{mod,interp}(f)$ is a linear interpolation of $\vec{\Phi}_\mathrm{mod}^{(a)}$ in the frequency grid.
Here, $\pi/10$ is chosen as the threshold to ensure the discretization error of the output is $\ll\pi$.

The modified gravity dataset will be split into a training set and a validation set. The training set, as its name indicates, will be used for the training of the VAE and the secondary networks. The validation set will be used to check that the trained networks can efficiently capture signals that the networks were not trained on. In this paper, 88\% of the modified gravity dataset will be used for training, while 12\% will be used for validation. 

In the following subsections, we will describe the behavior of the VAE and the secondary network, assuming that they work on elements of the modified gravity dataset, given by $\Xi^{(a)}_\alpha$ and $\vec{\Phi}_\mathrm{mod}^{(a)}$. 
However, one should keep in mind that these descriptions should also formally apply to any generic source parameters $\Xi_\alpha$ and phase modification function $\Phi_\mathrm{mod}(f)$. For the latter, a vector $\vec{\Phi}_\mathrm{mod}$ will be created following the procedure in Eq.~\eqref{eqn:npe_discretize}.

\subsection{The VAE network for the shape function} \label{sec:shape_vae}
As we have discussed in Sec.~\ref{sec:npe_sketch}, we will use a VAE to learn the shape function $S(\bar{f};n_i)$. A formal description is as follows. First, the encoder extracts information from the modified gravity data in a scale-invariant way:
\begin{align}
    \mu_i^{(a)}=\mu_i\big(\hat{\Phi}_\mathrm{mod}^{(a)}\big),\quad
    \sigma_i^{(a)}=\sigma_i\big(\hat{\Phi}_\mathrm{mod}^{(a)}\big), \label{eqn:encoder}
\end{align}
where $\hat{\Phi}_\mathrm{mod}^{(a)}=\vec{\Phi}_\mathrm{mod}^{(a)}\big/\big\lVert\vec{\Phi}_\mathrm{mod}^{(a)}\big\rVert$.
The decoder, then, takes a point $n_i^{(a)}$ on the latent unit hypersphere and maps it to a shape vector
\begin{align}
    \vec{S}^{(a)}=\vec{S}\big(n_i^{(a)}\big),
\end{align}
which is to be interpolated in the frequency grid to recover the continuous shape function of $\bar{f}$.

We connect the encoder and the decoder in the latent space, and the output of the decoder to the training data as follows. First, we require $\mu_i^{(a)}$ to be a unit vector, i.e.,~$\lVert\mu_i^{(a)}\rVert=1$. Second, we require that the random sample, taken from distributions associated with $\mu_i^{(a)}$ and $\sigma_i^{(a)}$, be generated from
\begin{align}
    n_i^{(a)}=z_i^{(a)}\big/\big\lVert z_i^{(a)}\big\rVert, \quad z_i^{(a)}\sim\mathcal{N}\big(\mu_i^{(a)},\sigma_i^{(a)}\big). \label{eqn:npe_variation}
\end{align}
Finally, the reconstruction loss is penalized in a scale-invariant way, namely
\begin{align}
    L_\mathrm{recon}=&\sum_{a} \Big\lVert\hat{\Phi}_\mathrm{mod}^{(a)}-\hat{S}\big(n_i^{(a)}\big)\Big\rVert^2\,. \label{eqn:npe_loss_vae}
\end{align}
Note here that the loss function depends on $\hat{S}^{(a)} = \vec{S}^{(a)}\big/\big\lVert\vec{S}^{(a)}\big\rVert$ and not on $\vec{S}^{(a)}$ alone, which ensures the same scale invariance as for $\hat{\Phi}_\mathrm{mod}^{(a)}$.

The encoder and the decoder will be neural networks, but before we define them, let us first discuss some npE specific choices for the network designs. 
First, because $\Phi_\mathrm{ppE}$ is modeled by two non-GR parameters $(b_\mathrm{ppE},\beta_\mathrm{ppE})$, we choose the VAE latent space to also be two-dimensional, which we find to be sufficient to faithfully represent these ppE theories. 
Based on that, we also require the two standard deviations of the encoder to be degenerate, i.e.~$\sigma_1=\sigma_2$. This logically makes sense because the true VAE parametrization only takes place on the unit circle, which is just 1-dimensional.

We require the shape function to take the form of a psuedo-PN expansion:
\begin{align}
    S(\bar{f};n_i)=\sum_{n=1}^{N_p} U_n(n_i)\,\bar{f}^{V_n(n_i)}, \label{eqn:npe_pseudo_pn}
\end{align}
where $U_n$ and $V_n$ are to be implemented as subnetworks of the decoder. This choice makes our shape network physically informed, allowing it to emulate any power-law behavior in the training set and recover the standard ppE modifications, while a generic network cannot. Equation~\eqref{eqn:npe_pseudo_pn} also formally addresses how the decoder output is interpolated in the frequency grid. The number of pseudo-PN terms $N_p$ is a hyperparameter of the VAE. 
When $N_p=1$, Eq.~\eqref{eqn:npe_pseudo_pn} resembles a na\"ive extension of the ppE framework, i.e.~upgrading the $b_\mathrm{ppE}$ from an integer to a real number. 
When $N_p\geq2$, Eq.~\eqref{eqn:npe_pseudo_pn} resembles an advanced ppE framework with higher-PN corrections included. In this case, we expect the VAE to learn a latent space with fewer dimensions as compared to the total number of PN coefficients to be otherwise included, and distribute a different mixture of PN orders with appropriate prior weights. In this paper, we will fix $N_p = 2$, but other choices can be studied in the future. 
Note that, usually, the decoder and the encoder are designed to mirror each other, but here we do not require this property. In particular, we do not ask the encoder to know anything about the pseudo-PN expansion of the decoder, because we wish the encoder to be able to handle more generic modifications. 

Let us now specify the lower-level implementation of the VAE.
We introduce three neural mappings, ${\mathcal{E}}\!\!: \,\mathbb{R}^{640}\mapsto\mathbb{R}^{3}$, $\mathcal{D}_U\!\!:\,\mathbb{R}^{2}\mapsto\mathbb{R}^{2}$ and $\mathcal{D}_V\!\!:\,\mathbb{R}^{2}\mapsto\mathbb{R}^{2}$. 
The neural map ${\mathcal{E}}$ takes in 640 real numbers associated with the modified phase at each frequency grid point, and it returns 3 real numbers associated with the 2 means of the latent space and the 1 standard deviation. The neural map $\mathcal{D}_U$ and $\mathcal{D}_V$, then, take 2 real numbers in the 2-D latent space and outputs two real numbers to be used as amplitudes and exponents in the pseudo-PN representation of the shape function. 
Each of these neural maps contains 4 fully connected hidden layers, and each hidden layer contains 512 neurons taking the ReLU activation\footnote{Here, ReLU stands for the rectified linear unit, whose functional form is ${\rm ReLU}(x)=x\,\Theta(x)$ where $\Theta(\cdot)$ is the Heaviside step function. Other example activation functions include the hyperbolic tangent function, the sigmoid function, etc. The ReLU is the standard choice for deep learning because it mimics biological behavior in the cortex~\cite{hahnloser2000digital}, while also addressing an issue common to neural networks, where numerical precision issues cause gradients to ``vanish,'' preventing further optimization~\cite{Goodfellow-et-al-2016}}. Therefore, the entire encoder-decoder network contains 3,486,727 parameters that we must train by minimizing the loss function.   

With this in hand, we implement the encoder and the decoder as follows. Consider the encoder first. For every element ${(a)}$ in the training set, 
\begin{align}
  \mu_i^{(a)} =& \frac{\mathcal{E}_i\big(\hat{\Phi}_\mathrm{mod}^{(a)}\big)-\mathcal{E}_i\big(-\hat{\Phi}_\mathrm{mod}^{(a)}\big)} {\sqrt{\sum_{j=1,2}\left[\mathcal{E}_j\big(\hat{\Phi}_\mathrm{mod}^{(a)}\big)-\mathcal{E}_j\big(-\hat{\Phi}_\mathrm{mod}^{(a)}\big)\right]^2}},\label{eqn:impl_enc_mu} \\
    \sigma_i^{(a)} =& \exp\left[ \mathcal{E}_3\big(\hat{\Phi}_\mathrm{mod}^{(a)}\big) + \mathcal{E}_3\big(-\hat{\Phi}_\mathrm{mod}^{(a)}\big) \right], \label{eqn:impl_enc_sigma} 
\end{align}
where the $i$, $j$ and $3$ subscripts in $\mathcal{E}$ indexes over the 3 real numbers in the output of $\mathcal{E}$ (the same notation also applies to other neural maps below). Note that both $\sigma_{1}$ and $\sigma_{2}$ comes from the same output $\mathcal{E}_3$, as previously stated.
We implement the decoder as
\begin{align}
    U_n^{(a)} =& \mathcal{D}_{U,n}\big(n_i^{(a)}\big) - \mathcal{D}_{U,n}\big(-n_i^{(a)}\big),\label{eqn:impl_dec_u}\\
    V_n^{(a)} =& \mathcal{D}_{V,n}\big(n_i^{(a)}\big) + \mathcal{D}_{V,n}\big(-n_i^{(a)}\big),\label{eqn:impl_dec_v}
\end{align}
The normalization in Eq.~\eqref{eqn:impl_enc_mu} ensures that $\mu_i$ is on the unit circle. The exponential in Eq.~\eqref{eqn:impl_enc_sigma} ensures that $\sigma_i$ is positive. The symmetrization and antisymmetrization through Eqs.~\eqref{eqn:impl_enc_mu}--\eqref{eqn:impl_dec_v} are in response to the requirement that $S(\bar{f})$ be modeled as an odd function of $n_i$. 

To summarize, the elemental components of the encoder and the decoder are presented in Fig.~\ref{fig:npe_networks} (a) and (b), respectively. These two networks compose the shape VAE, and the way they connect for training follows Fig.~\ref{fig:npe_connect} (b). 
The training process minimizes the VAE loss function in Eq.~\eqref{eqn:vae_loss_template}, in which the reconstruction loss is further specified by Eq.~\eqref{eqn:npe_loss_vae}. 
We customize the VAE with additional operations (uncolored rounded rectangles in the figures) beyond the standard formulation in Fig.~\ref{fig:vae}, including normalization, symmetrization, and especially the pseudo-PN expansion. 

\subsection{The secondary network for the scale function} \label{sec:secondary_network}
Let us now describe the secondary network for the scale function $T(\Xi_\alpha,n_i)$. 
Unlike the modeling of $S(\bar{f})$, the accuracy of $T$ does not significantly impact the npE tests of GR. More precisely, a mismodeling of $T$ changes the EFT prior boundary, which only affects the magnitude of the largest modifications that will be covered by the latent space. 
In our case, the EFT prior boundary is determined by the map to the largest ppE modifications we allow, characterized by Eq.~\eqref{eqn:npe_max_beta}. In this way, if GR is modified through a small deformation, the resulting GW signals should never map to areas of the latent space that are anywhere close to our EFT prior boundary. 

With that in mind, we follow a simple design for the secondary network. We introduce one neural mapping, ${\mathcal{T}}\!\!: \,\mathbb{R}^{6}\mapsto\mathbb{R}$, which takes in 6 real numbers associated with the 4 source parameters plus 2 latent space coordinates, and returns 1 real number associated with the scale value. 
As with the VAE neural maps, we also implement $\mathcal{T}$ with 4 fully connected hidden layers, each containing 512 neurons taking the ReLU activation. The entire secondary network contains 1,054,721 parameters for training. 

The scale function is then constructed with the same strategy presented in Eq.~\eqref{eqn:impl_enc_sigma}:
\begin{align}
    T^{(a)} =& \exp\left[ \mathcal{T}\big(\Xi_\alpha^{\prime(a)},n_i^{(a)}\big) + \mathcal{T}\big(\Xi_\alpha^{\prime(a)},-n_i^{(a)}\big) \right], \label{eqn:impl_2nd}
\end{align}
where the exponential ensures that $T$ is positive, and the symmetrization ensures that $T$ is an even function of $n_i$. 
Before inserting the source parameters $\Xi_\alpha$ in $\mathcal{T}$, we change coordinates in the source parameter space to
\begin{align}
    \Xi_\alpha'=\left(\ln\mathcal{M},q,\frac{\chi_1+\chi_2}{2},\frac{\chi_1-\chi_2}{2}\right),
\end{align}
which are better adapted for the exploration of the likelihood in Bayesian parameter estimation of GW signals from BBH systems.
The above specifications of the secondary network is summarized in Fig.~\ref{fig:npe_networks} (c).

To train the secondary network, we connect it to the shape VAE as shown in Fig.~\ref{fig:npe_connect} (c) and choose the following loss function:
\begin{align}
    L_\mathrm{2nd}=&\sum_{a} \left[\ln \frac{\big\lVert\vec{\Phi}_\mathrm{mod}^{(a)}\big\rVert}{T\big(\Xi_\alpha^{(a)},\mu_i^{(a)}\big)\,\big\lVert\vec{S}\big(\mu_i^{(a)}\big)\big\rVert} \right]^2, \label{eqn:npe_loss_2nd}
\end{align}
% \dc{why not use the notation $L_\mathrm{\mathcal{T}}$ since the network is called $\mathcal{T}$.}
% \yx{I feel that $L_\mathrm{\mathcal{T}}$ does not sound consistent with $L_\mathrm{VAE}$. The latter is brought up early in Eq.~\eqref{eqn:vae_loss_template} and applys to something that not only employs the lower-level $\mathcal{E}$ and $\mathcal{D}$, but also has the pseudo-PN expansion, symmetrization, and normalization. Also we've been saying ``the VAE network'' and ``the secondary network'' as if they are at the same level. So I'd stick to the word ``secondary'', or ``2nd'' here.}
% \gn{Wouldn't this be better as $L_{recon} = L_{shape} + L_{scale}$ rather than $L_{VAE}$ and $L_{2nd}$? - i.e. what the Loss functions are doing, rather than how they are implemented.}
% \yx{That makes sense, but I borrowed $L_\mathrm{VAE}$ from Eq.~\eqref{eqn:vae_loss_template}, so I feel more coherent to say $L_\mathrm{2nd}$. I think either way is clear. Let me think more carefully...}
where $\mu_i$ and $\vec{S}$ are the encoder mean and the decoder reconstruction from the shape VAE. Because the shape VAE should have been trained before this stage, $\mu_i$ and $\vec{S}$ should be known functions, and all neural parameters in $\mathcal{E}$, $\mathcal{D}_U$ and $\mathcal{D}_V$ should be kept frozen when training $L_\mathrm{2nd}$. 

The above loss function is designed to penalize the fractional error in the modeling of the scale function. 
Because $\mu_i^{(a)}$ is on the unit circle, the denominator of Eq.~\eqref{eqn:npe_loss_2nd} essentially gives $\big\lVert\vec{\Phi}_\mathrm{npE}(\Xi_\alpha^{(a)},\mu_i^{(a)})\big\rVert$, or the norm of the npE phase at the EFT prior boundary, where the training data is defined (see Sec.~\ref{sec:data_set}). Assuming that $\vec{S}\big(\mu_i^{(a)}\big)$ has learned the shape of $\vec{\Phi}_\mathrm{mod}^{(a)}$, any difference between $\big\lVert\vec{\Phi}_\mathrm{npE}(\Xi_\alpha^{(a)},\mu_i^{(a)})\big\rVert$ and $\big\lVert\vec{\Phi}_\mathrm{mod}^{(a)}\big\rVert$ can only be attributed to modeling error in $T$, which is summed over in Eq.~\eqref{eqn:npe_loss_2nd} in terms of a fractional error.

\subsection{Training procedure and results} \label{sec:train}
We implement the networks described in Secs.~\ref{sec:shape_vae} and \ref{sec:secondary_network} with \texttt{PyTorch}~\cite{1912.01703}. 
For training, we choose a combination of the \texttt{AdamW} optimizer~\cite{1711.05101}, with a weight decay of $10^{-4}$, and the \texttt{ExponentialLR} scheduler, with a learning rate decay of by a factor of $0.9$ every epoch\footnote{In each epoch, the training set is randomly split into mini-batches. Then minimization of the loss is progressively performed by the optimizer batch by batch so as to evade local minimums. The completion of an epoch marks one full iteration over the entire training set. The learning rate controls how much the network parameters get updated by each minibatch of training data. At the end of each epoch, the learning rate is reduced by the scheduler, so that optimization converges over epochs. See~\cite{lu2024gradient} for more details about the related concepts.}. 
We start by training the VAE for $50$ epochs with a batch size of $64$ and an initial learning rate of $10^{-4}$. The coefficient of the KL loss is fixed to $\kappa=10^{-6}$ during the process. Then, we freeze the VAE parameters and train the secondary network for another 50 epochs using the same batch size and initial learning rate.
Here, the value of $\kappa$ is empirically determined to give the best learning result for the shape function. The other settings, such as the learning rate and the batch size, are free to vary in some reasonable range, at most impacting the speed of convergence.

After training is complete, we can then use the decoder and the secondary network as a model for the modified waveform phase. 
Although the neural network infrastructure may slow down the evaluation of the npE waveform, we have checked that this is not a significant effect for the parameter estimation studies we considered, at most doubling the waveform evaluation time in single-CPU test runs (for more detail, see Appendix~\ref{apd:npe_time_complexity}). 

Figure~\ref{fig:npe_training_history} shows the training and validation losses as functions of epochs for the VAE [panel (a)] and the secondary network [panel (b)]. 
Observe that both the VAE and the secondary network reach convergence within the maximum number of epochs explored. 
The validation loss of the secondary network is considerably higher than the training loss after the loss has converged, implying overfitting. However, as mentioned in Sec.~\ref{sec:secondary_network}, the secondary network does not necessarily require high accuracy, since it only affects the EFT prior boundary. 
The validation loss after convergence implies a $\sim6\%$ fractional error for the modeling of the scale function $T$\footnote{According to Eq.~\eqref{eqn:npe_loss_2nd}, this error is roughly the exponential of the square-root average loss minus one. Here, the average loss is read from the rightmost orange marker in Fig.~\ref{fig:npe_training_history} (b).}, which is acceptable for our purposes. 
\begin{figure}
    \centering
    \includegraphics[width=0.48\textwidth]{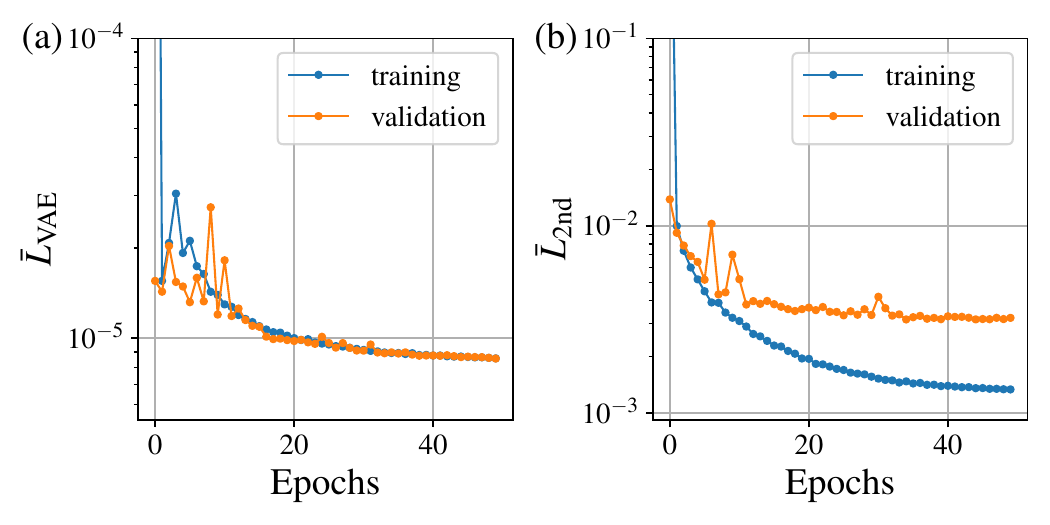}
    \caption{
    % \gn{They already share an axis to suggest just combining the two into a single figure, because even if you separate the two components and train them sequentially, the thing that matters is the full npE.}
    % \yx{Let me try and see}
    Training history for (a) the VAE and (b) the secondary network of the npE framework. The history is given as the training and validation losses (each normalized by the size of the corresponding dataset) versus the number of training epochs. Observe that both networks reach convergence within the maximum number of epochs explored.}
    \label{fig:npe_training_history}
\end{figure}

\begin{figure}[htbp]
    \centering
    \includegraphics[width=0.48\textwidth]{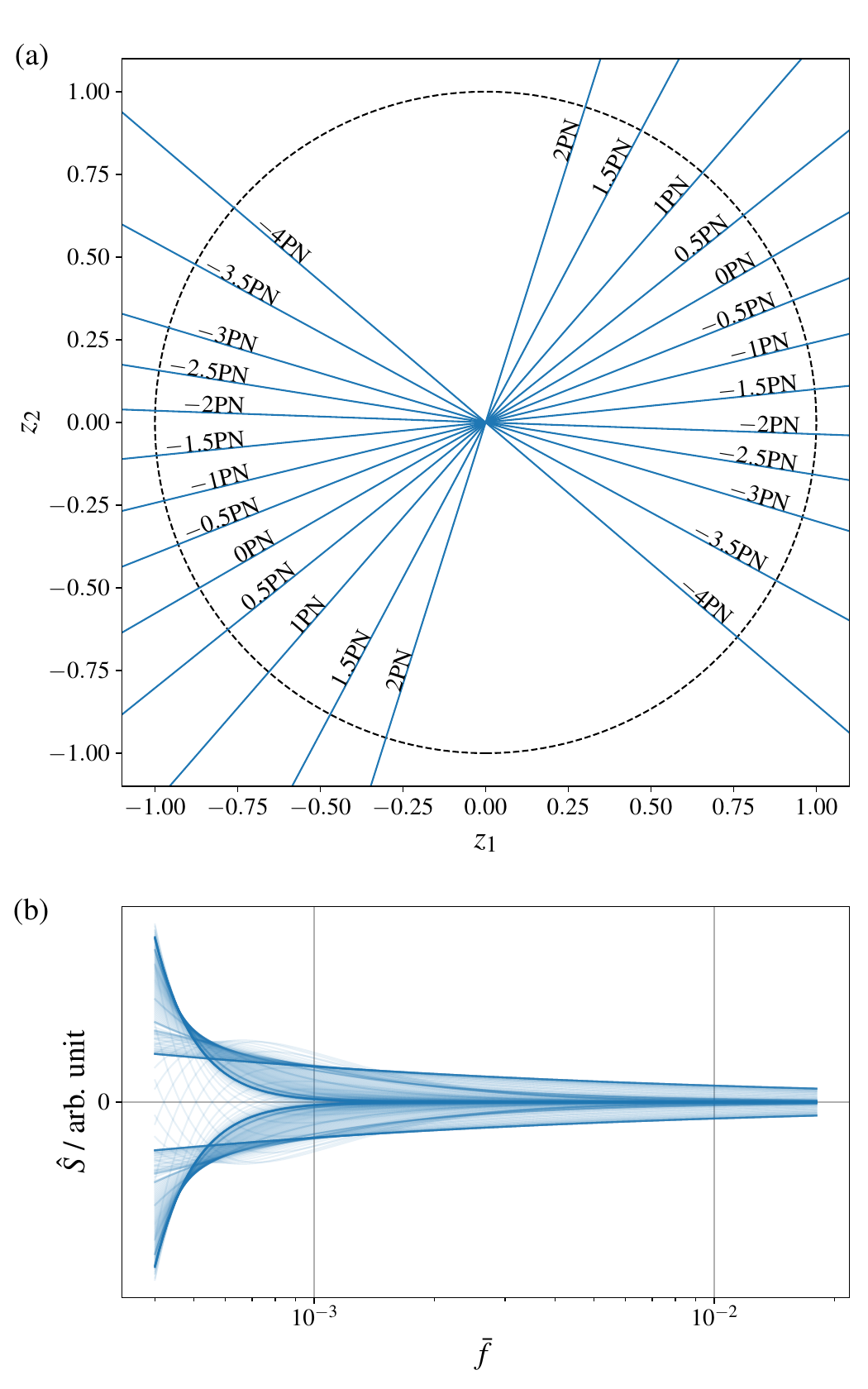}
    \caption{
    Latent space angular parametrization learned by the VAE. (a) shows the representation of ppE theories (given by the encoder mean) in the latent space. Observe that each ppE index maps to a radial line. Different indices are separated, and are put in order in the latent space. (b) overlays the shape functions generated from a grid of the latent space polar angles separated by $1^\circ$. Observe that these shape functions form a continuous band, confirming that the learned npE angular parametrization is continuous.}
    \label{fig:npe_representation}
\end{figure}

The performance of the npE is determined by the accuracy of the reconstruction of the shape function, which in turn depends on how well the VAE learned the shape of the training data [see Appendix~\ref{apd:npe_reconstruction} for examples using the npE framework to reproduce the ppE modeling of EdGB theory and dynamical Chern-Simons (dCS) gravity.] However, to assess whether the npE is appropriate for tests of GR, we must carry out Bayesian parameter estimation. The key question is how much information we can get from the posterior when using the VAE and the latent space to parametrize the waveform, which we leave for discussion in Sec.~\ref{sec:test_npe}.
For the moment, however, let us first take a look at how the VAE perceives and extends the ppE theories in the modified gravity dataset, which is summarized in Fig.~\ref{fig:npe_representation}. 

The representation of ppE theories in the latent space (given by the encoder mean) is shown in Fig.~\ref{fig:npe_representation} (a). 
These ppE theories map to radial lines, which is not surprising given the npE prescription in Eqs.~\eqref{eqn:npe_decomp} and \eqref{eqn:npe_parity}, and the fact that ppE phases associated with the same $b_\mathrm{ppE}$, do not vary in shape. 
What matters is that these ppE theories are well organized in the latent space. In particular, the mapping is \textit{injective}, in the sense that there is a one-to-one relation between the ppE index $b_\mathrm{ppE}$ (or PN order) and the polar angle (up to $\pi$) of the latent line, so each theory maps to one line and different theories separate. Moreover, the mapping is \textit{ordered}, in the sense that theories with deformations that enter at similar PN orders are also close to each other in angle in the latent space. 
We highlight that we arrive at this result in a completely unsupervised manner, i.e.,~ordering is not imposed explicitly.

Figure~\ref{fig:npe_representation} (a) also presents ppE theories that introduce modifications that enter at half-integer PN orders. These theories are not included in the modified gravity dataset for training, but they are mapped to the latent space in a similar way to those in the training set. This means that the VAE understands well the structure of ppE theories in the latent space. 
These observations are further confirmed by Fig.~\ref{fig:npe_representation} (b), where we present the shape functions generated from a sample of latent space angles, which is much denser than the distribution of angles of the ppE theories contained in the training set. These functions form a band when overlaid, suggesting that the ppE parametrization has been continuously extended by the npE latent space. 

\begin{figure}[htbp]
    \centering
    \includegraphics[width=0.48\textwidth]{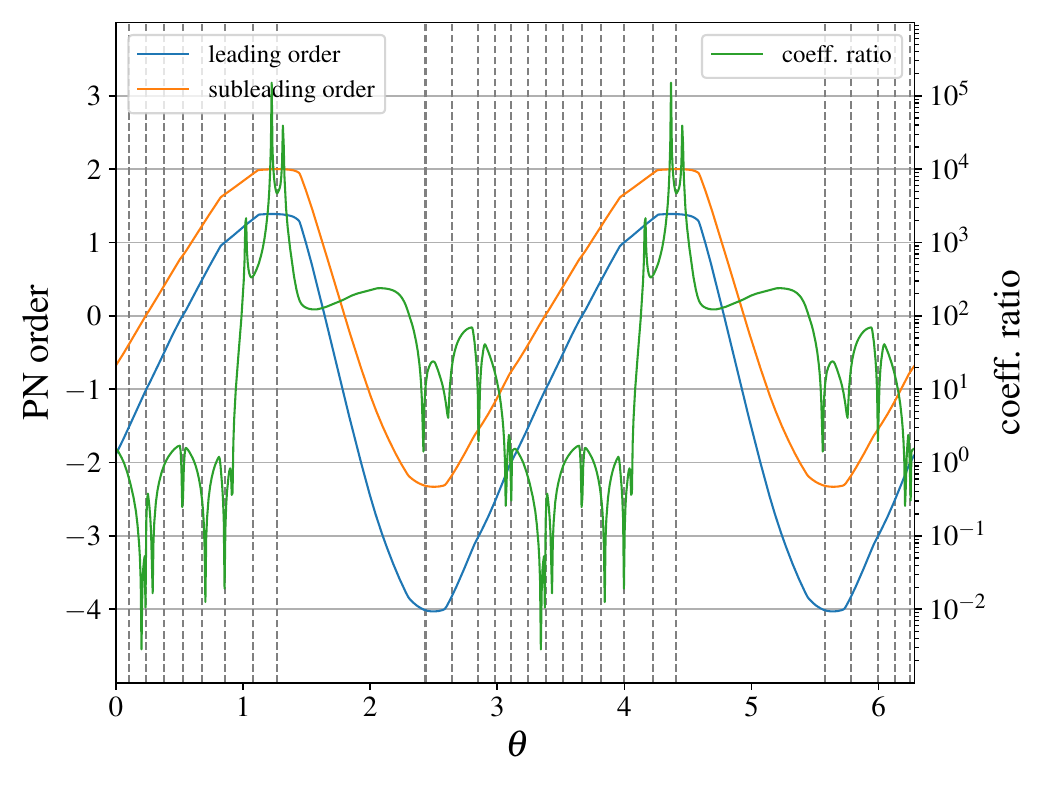}
    \caption{Pseudo-PN expansion across the polar angle $\theta$ in the latent space learned by the VAE. The blue (orange) curve shows the effective PN order of the leading (subleading) pseudo-PN term, which refers to the left y-axis. The green curve shows the ratio between the two effective PN coefficients, with the subleading one divided by the leading one, which refers to the right y-axis. 
    The gray dashed lines correspond to the angles of those ppE theories shown in Fig.~\ref{fig:npe_representation} (a).
    Observe that blue curve and the orange curve do not cross, and the coefficient ratio is $\gtrsim1$ for most of the polar angles between the ppE lines, which means that the two terms in the expansion are not degenerate. 
    Also the coefficient ratio is $\gg1$ approaching the gaps between ppE-dense regions, suggesting that non-PN behavior may be included in these gaps.}
    \label{fig:npe_pseudopn}
\end{figure}

As discussed in Sec.~\ref{sec:shape_vae}, the pseudo-PN expansion adopted by the decoder leads to a trivial extension of the ppE parametrization when $N_p=1$. Although we have chosen $N_p=2$, the two terms in Eq.~\eqref{eqn:npe_pseudo_pn} may lead to degeneracies after training, and reduce to the trivial case. Therefore, we must investigate the behavior of the learned pseudo-PN expansion across the latent space, which is illustrated in Fig.~\ref{fig:npe_pseudopn}. 
In this figure, the blue (orange) curve shows the effective PN order of the leading (subleading) pseudo-PN term, i.e.~the term with a smaller (greater) $V$ exponent on $\bar{f}$ in Eq.~\eqref{eqn:npe_pseudo_pn}, with respect to the polar angle of the latent space $\theta$. 
The green curve then shows the ratio between the two effective PN coefficients, with the subleading one divided by the leading one. 
Observe that the subleading PN order never coincides with the leading PN order, i.e.~the blue and the orange curves never cross. Observe also that the coefficient ratio is $\gtrsim1$ (right y-axis) for most of the polar angles between the ppE lines. 
These two observations suggest that the double-term pseudo-PN expansion learned by our VAE does not reduce to the single-term case. Thus, we verify that we have obtained a npE parametrization that nontrivially extends the ppE framework. 

Another observation one can make about Fig.~\ref{fig:npe_pseudopn} is that the coefficient ratio is smaller at the angles where the ppE lines are denser, and it is greater at the angles where the ppE lines are sparse, i.e.~within the ``big gap'' between $-4$PN to $2$PN order. 
This suggests that the ppE-denser region is more ppE-like, with a better-ordered pseudo-PN expansion. On the other hand, the ppE-sparse region is where the decoder behavior is not of ppE type. 
One may then define a reference angle $\theta_\mathrm{ref}$ inside the ppE-sparse region, and define the bilateral deviation $z_b$ and the theory angle $\varphi$ as
\begin{align}
    z_b =\,& \lVert z\rVert\, \mathrm{sign}\left[\sin(\theta-\theta_\mathrm{ref})\right], \\
    \varphi =\,& \mathrm{mod}(\theta-\theta_\mathrm{ref},\, \pi).
\end{align}
The reference angle $\theta_\mathrm{ref}$ can be chosen as the angle at which the shape function varies the most rapidly, which for us is at $\theta_\mathrm{ref}=1.88$ rads, with details provided in Appendix~\ref{apd:angle_ref}.
Under this description, each ppE index $b_\mathrm{ppE}$ is represented by a unique $\varphi$ value, and $z_b$ scales linearly with $\beta_\mathrm{ppE}$, with both positive and negative values allowed. 
In the next section, we will find this $z_b$-$\varphi$ parametrization convenient for comparison between npE and ppE Bayesian parameter estimation.

\section{Testing GR with the npE Framework} \label{sec:test_npe}
In this section, we investigate the practical use of the npE framework for testing GR, by running Bayesian parameter estimation with a npE-parametrized waveform template on simulated signals with or without modifications to GR.
The discussion will be carried out under the following four different studies:
\begin{enumerate}
    \item \textit{Constraining deviations from GR.} We inject a GR signal and attempt to recover it with an npE model and a ppE model to find bounds on non-GR modifications. 
    \item \textit{Detecting leading-PN-order deviations from GR}. We inject a ppE signal and attempt to recover it with both an npE model and a ppE model to allow for comparisons. 
    \item \textit{Detecting high-PN-order deviations from GR}. We inject a ppE-like signal with high-PN-order corrections to the leading-order ppE deviation, and attempt to recover it with an npE model.
    \item \textit{Detecting non-PN-like deviations from GR}. We inject a non-PN-expandable signal and attempt to extract it with an npE model. 
\end{enumerate}
Each of these studies will be elaborated in the next subsections, once we have provided a description on the general setup of the injection-recovery runs.

\subsection{Injection-recovery setup} \label{sec:test_setup}
All injected signals will consist of GWs generated by quasicircular, nonspinning BBH systems. 
In particular, we will consider two prototypical systems: a heavy binary with total mass $M_\mathrm{inj}=35\,M_\odot$, and a lighter binary with $M_\mathrm{inj}=15\,M_\odot$, both with a mass ratio of $q_\mathrm{inj}=2/3$. 
These signals will be assumed to have been detected by a Hanford-Livingston-Virgo network, with an O5 design sensitivity~\cite{KAGRA:2013rdx}\footnote{\url{https://dcc.ligo.org/LIGO-T2000012/public}}.
We do not inject the signals into specific realizations of noise because we wish to study the averaged outcome of parameter estimation, which is independent of a given noise artifact.
The signals are analyzed in a frequency range between $10\,\mathrm{Hz}$ and $0.018/M_\mathrm{inj}$, the latter of which is chosen so that only the inspiral part of the signal is taken into account. 
For each signal, we scale the luminosity distance such that the network SNR is always fixed at $40$ for the frequency range specified above. 

The signals and models we consider here are constructed as follows. All GR signals are built directly from the IMRPhenomD waveform model $\tilde{h}_\mathrm{GR}$. Non-GR signals are constructed as deformations of the IMRPhenomD model, either through leading-PN-order (ppE) deviations, high-PN-order (ppE-like) deviations, or non-PN-like deviations of the waveform phase. These modifications $\Phi_\mathrm{mod}$ are added to the IMRPhenomD phase in the inspiral region, as explained in Eq.~\eqref{eqn:npe_generic_mod}, up to the inspiral frequency cutoff at $Mf=0.018$. Therefore, the injected signals do not contain any information in the intermediate or merger-ringdown regimes ($Mf>0.018$). The non-GR models we use to recover these signals are built in exactly the same way, with the exception that they are allowed to include the intermediate and merger-ringdown regimes (these regimes will be needed in case the sampler explores a region of parameter space where $M>M_\mathrm{inj}$.) We do not introduce explicit non-GR effects in these regimes, although some do trickle down from the inspiral regime through the continuity and differentiability matching conditions at the interphases.

The injected signals are then recovered using Bayesian statistical theory, using \texttt{Bilby}~\cite{Ashton:2018jfp} and the likelihood nested sampler \texttt{dynesty}~\cite{1904.02180} to perform parameter estimation. 
For simplicity, we only sample those model parameters that enter the GW phase of the recovery model, including the masses, the spins, the time and phase of coalescence, and any modified-gravity parameters. The other parameters, such as the distance and the sky location, are fixed to the injected values. 
Sampling over all amplitude and phase parameters simultaneously would increase the computational cost and the correlations in the posterior, which we leave for future studies.

The prior for sampling the parameters in the phase of the recovery model are chosen as follows. 
The prior for the masses is uniform in $\mathcal{M}_c$ and $q$. The range of $\mathcal{M}_c$ is $[10,20]\,M_\odot$ for the heavier BBH and $[5,8]\,M_\odot$ for the lighter BBH, whereas the range of $q$ is always $[0.125,1]$. 
The prior for each dimensionless spin $\chi_{1,2}\in[-1,1]$ is proportional to $\ln|\chi_{1,2}|$, which can be derived as the aligned-spin counterpart of the precessing, uniform spin magnitude prior by marginalizing the latter over all angular degrees of freedom~\cite{Lange:2018pyp}. 
The phase and time of coalescence each take a uniform prior, with the former bound by $[0,2\pi]$, and the latter bound by a $\pm0.1\,\mathrm{s}$ range around the injected value. For the ppE recovery, we choose the prior for $\beta_\mathrm{ppE}$ to be uniform in a sufficiently wide range, $[-100,100]\,\beta_\mathrm{ppE,max,inj}$, where $\beta_\mathrm{ppE,max,inj}$ is Eq.~\eqref{eqn:npe_max_beta} evaluated with the injected source parameters and $b_\mathrm{ppE,inj}$.

The prior choice for the npE recovery raises some complexity. Following the discussion of Sec.~\ref{sec:npe_sketch}, $(z_1,z_2)$ should take a ``flat prior'' in the unit circle.
The most straightforward implementation is a uniform prior in $(z_1,z_2)$, but doing so leads to an unsatisfactory prior in the magnitude $\lVert z_i\rVert$, in particular excluding GR due to the divergence of the Jacobian at $z_i=0$. A uniform prior in $(z_1,z_2)$ would therefore artificially disfavor GR in the marginalized posterior of $\lVert z_i\rVert$. For this reason, we will also consider a prior that is uniform in $\lVert z_i\rVert$ and $\theta$, or equivalently, uniform in $z_b$ and $\varphi$. We will present results obtained choosing both types of priors in the following subsections. 

\subsection{Constraining deviations from GR} \label{sec:constrain_gr}

\begin{figure*}
    \centering
    \includegraphics[width=0.98\textwidth]{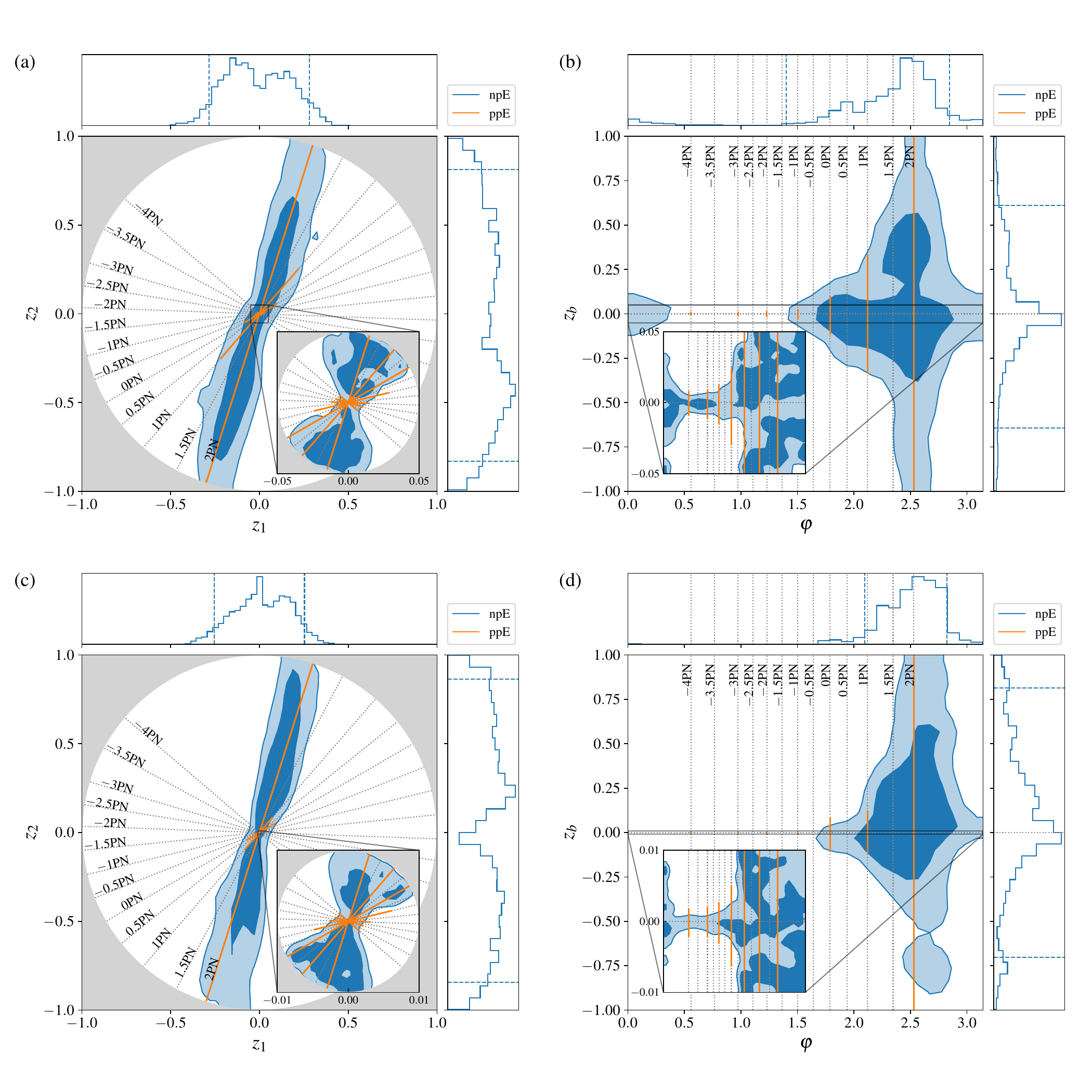}
    \caption{Predicted constraints on non-GR modifications with the npE framework. 
    GR injections of a heavy BBH [panels (a) and (b)] and a light BBH [panels (c) and (d)] are recovered with an npE model in the $(z_1,z_2)$ parametrization [panels (a) and (c)] and the $(z_b,\varphi)$ parametrization [panels (b) and (d)]. The posterior distributions are shown in blue, where the light and dark blue boundaries represent a 90\% and 50\% credible contours respectively. For reference, ppE models (i.e.,~with fixed ppE indices) are represented in this figure with gray dotted lines, while regions in the parameter space that are outside of the npE prior are shaded in gray. In order to allow for comparisons between the npE analysis and an independent ppE analysis, we also present the 90\% credible region of the $\beta_\mathrm{ppE}$ posterior, mapped to the $(z_1,z_2)$ or $(z_b,\varphi)$ latent parameter space with solid orange lines.
    The insets in each panel shows the npE recovery with a prior range that is smaller than the standard one given by the EFT prior boundary, which is indicated by the limit of the inset axis. 
    Each panel is also accompanied by 1D marginalized posteriors, where the dashed lines mark the 90\% credible intervals. 
    Observe that the npE model is able to constrain a wide range of the latent parameter space, with the constraints becoming more stringent in the direction of negative PN order deviations, i.e.~the white regions in the panels are excluded at 90\% confidence. Higher PN order deviations, however, are constrained less well, as expected. Finally, observe that the npE 90\% credible contours are comparable to the ppE 90\% credible intervals, when the prior range is properly chosen to resolve the npE contours near the origin. This indicates that the npE model is as good as the ppE one at constraining deviations from GR, with the advantage of being able to explore a wider region of theory parameter space. 
    }
    \label{fig:test_constrain_gr}
\end{figure*}

Let us begin by considering GR injections and an npE model for recovery. 
Figure~\ref{fig:test_constrain_gr} shows the npE posteriors in the $(z_1,z_2)$ and the $(z_b,\varphi)$ parametrizations for both the heavy and the light BBHs. For comparison, this figure also shows the ppE posteriors for a fixed set of ppE indices $b_\mathrm{ppE}$, which are the same as those used in the npE training set. For each fixed value of $b_\mathrm{ppE}$, the $\beta_\mathrm{ppE}$ posterior is transformed to the latent $(z_1,z_2)$ representation using the npE encoder to allow for direct comparisons.  

Let us first look at the joint posterior of npE parameters [$(z_1,z_2)$ or $(z_b,\varphi)$] in the main plot of each panel. Each point on the 90\% credible contour (CC) can be interpreted as a constraint on the corresponding type of modification. Thus, we confirm that the npE framework can be used to constrain modifications to GR.

One may observe, however, that the constraint is tighter for modifications at lower (and especially at negative) PN orders and for the source with smaller masses. The same pattern is also presented by the ppE 90\% credible intervals (CIs), which is expected because lower-PN corrections are detected with more effective cycles~\cite{Chamberlain:2017fjl,Sampson:2014qqa}.
For the low-mass binary, the difference is further magnified as more early-inspiral signal is captured in the detector band.
As a side effect, the marginalized posteriors of the npE deviation parameters $(z_1,z_2,z_b)$ fail to give any useful information and are simply overwhelmed by the unconstrained $2$PN modifications in the latent space.
Thus, the npE joint posterior is necessarily needed when constraining modifications to GR. 

We may further compare the constraining power of the npE 90\% CC to that of the ppE 90\% CIs. 
At first glance, the former is tangent to the latter, except for those negative-PN modifications whose ppE constraints are much tighter. 
However, we note that the npE 90\% CC at negative-PN angles are not sufficiently resolved in the main plots of Fig.~\ref{fig:test_constrain_gr}. In order to numerically generate the CC, the posterior sample has to be binned, and the resolution of the resulting CC is bottlenecked by the finite size of the bins. This is particularly an issue at those negative-PN angles, because the CC in that region is too narrow to cover a sufficient number of bins. 

We note that the number of bins is ultimately limited by the finite size of the posterior sample, so a solution is to ask the sampler to generate more sample during parameter estimation. However, this solution would not be computationally affordable, as we probably need to increase the sample size by an order of magnitude to see any significant improvement.
Here, we take a different route. Instead of enlarging the entire posterior sample, we reduce the EFT prior boundary and run parameter estimation one more time with the same posterior sample size. This effectively ``zooms'' into the previously under-resolved region, and we show the additional posterior with finer bins in the inset of each panel of Fig.~\ref{fig:test_constrain_gr}.
This allows us to see that the npE 90\% CC still follows the width of each ppE 90\% CI at negative-PN angles, and hence, we conclude that the npE 90\% CC is as powerful as the ppE 90\% CIs at constraining non-GR theories. 

\subsection{Detecting leading-PN-order deviations from GR} \label{sec:detect_ppe}
\begin{figure*}
    \centering
    \includegraphics[width=0.98\textwidth]{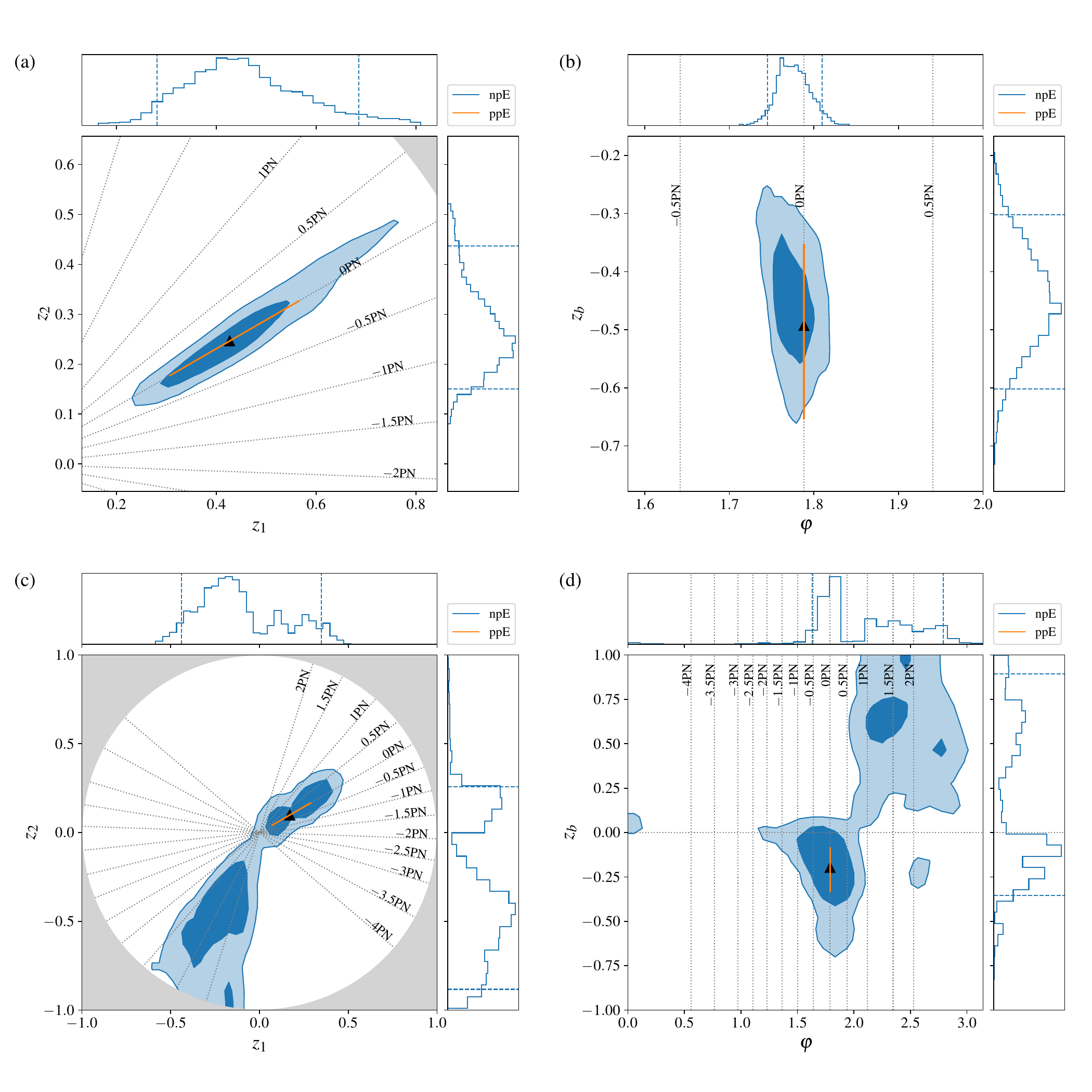}
    \caption{Posteriors recovering a ppE modification with the npE framework. 
    The injected signal is generated with the heavy BBH source in a theory that modifies the GR waveform only at the $0$PN order.
    The magnitude of modification chooses from $\beta_\mathrm{ppE,inj}/\beta_\mathrm{ppE,max}=50\%$ [panels (a) and (b)] and $\beta_\mathrm{ppE,inj}/\beta_\mathrm{ppE,max}=20\%$ [panels (c) and (d)], and the npE recovery is made with the $z_1$-$z_2$ parametrization [panels (a) and (c)] and the $z_b$-$\varphi$ parametrization [panels (b) and (d)].
    The plots follow the same format as in Fig.~\ref{fig:test_constrain_gr}. 
    In addition, an up-pointing triangle is added to each panel to mark the npE representation of the injected phase modification, using the npE encoder.
    Observe that, when $\beta_\mathrm{ppE,inj}$ is sufficiently large, the npE test detects the modification as effectively as the ppE one, plus that the npE posterior selects the correct type of theory. When $\beta_\mathrm{ppE,inj}$ is small, however, the npE test is less sensitive than the ppE one, with correlation built up in the posterior. 
    }
    \label{fig:test_deviation_examples}
\end{figure*}

Next, we consider injections with ppE modifications that only have a leading-PN-order term, and recover with an npE model. 
This way, we use the npE framework to detect non-GR modifications that are of the same type as those appearing in the npE training set.
Figure~\ref{fig:test_deviation_examples} shows an example, in which the injected signals are generated with $0$PN modifications at two different magnitudes. 
Again, we overlay the ppE results, but this time we only include the ppE posteriors obtained from the same ppE index as in the injection. 

Figure~\ref{fig:test_deviation_examples} allows us to make several observations. 
First, observe that for $\beta_\mathrm{ppE,inj}/\beta_\mathrm{ppE,max}=50\%$, the npE framework successfully detects the modification, because the 90\% CCs of the joint posteriors exclude GR at $z=0$. This is consistently suggested by the 90\% CIs of the npE deviation parameters $(z_1,z_2,z_b)$, and the results are comparable with the ppE posterior. 
Furthermore, the type of modification is also correctly resolved, as the 90\% CCs of the joint posterior and the 90\% CI of the theory angle $\varphi$ exclude all other PN orders. This \textit{outperforms} the ppE framework, which requires running parameter estimation multiple times and combining results from all possible ppE indices.
We also note that this is a case in which the $z_b$-$\varphi$ parametrization becomes useful and allows us straightforwardly read-off of the magnitude and type of the modification through the marginalized posteriors. 

For $\beta_\mathrm{ppE,inj}/\beta_\mathrm{ppE,max}=20\%$, however, the npE posterior fails to detect the modification, either with the 90\% CCs of the joint posteriors or with the 90\% CIs of the marginalized posteriors. This is a case in which the ppE framework is better, as the ppE 90\% CI still successfully excludes GR. 
The underlying difference is that the npE latent space contains multiple theories that may correlate with each other and worsen the posterior. Such ambiguity between theories is activated when the injected modification is small and everything returns to GR. This explains why the same issue does not occur in the previous case with a larger $\beta_\mathrm{ppE,inj}$.

Given the findings above, it may be interesting to find the critical level of modification which, when used for injection, allows the npE test to detect the existence, or resolve the type, of the modification. 
For convenience, let us refer to these critical values as the detection boundary and the resolution boundary, respectively. 
We carry out the following procedure to determine these boundaries:
\begin{enumerate}
    \item For each $b_\mathrm{ppE}$, begin with $\beta_\mathrm{ppE,inj}/\beta_\mathrm{ppE,max}=100\%$ and run the npE test. 
    \item If the result suggests detection or resolution, reduce $\beta_\mathrm{ppE,inj}$ by a factor of 2 and rerun the npE test. 
    \item Repeat step 2 until the detection or resolution is lost, and obtain an interval within which the corresponding boundary lies. 
\end{enumerate}
In this procedure, each npE test is done with the $z_b$-$\varphi$ parametrization. A detection is claimed when the 90\% CI of $z_b$ excludes GR, and a resolution is claimed when 90\% CI of $\varphi$ excludes angles of any other PN orders. 

\begin{figure}
    \centering
    \includegraphics[width=0.48\textwidth]{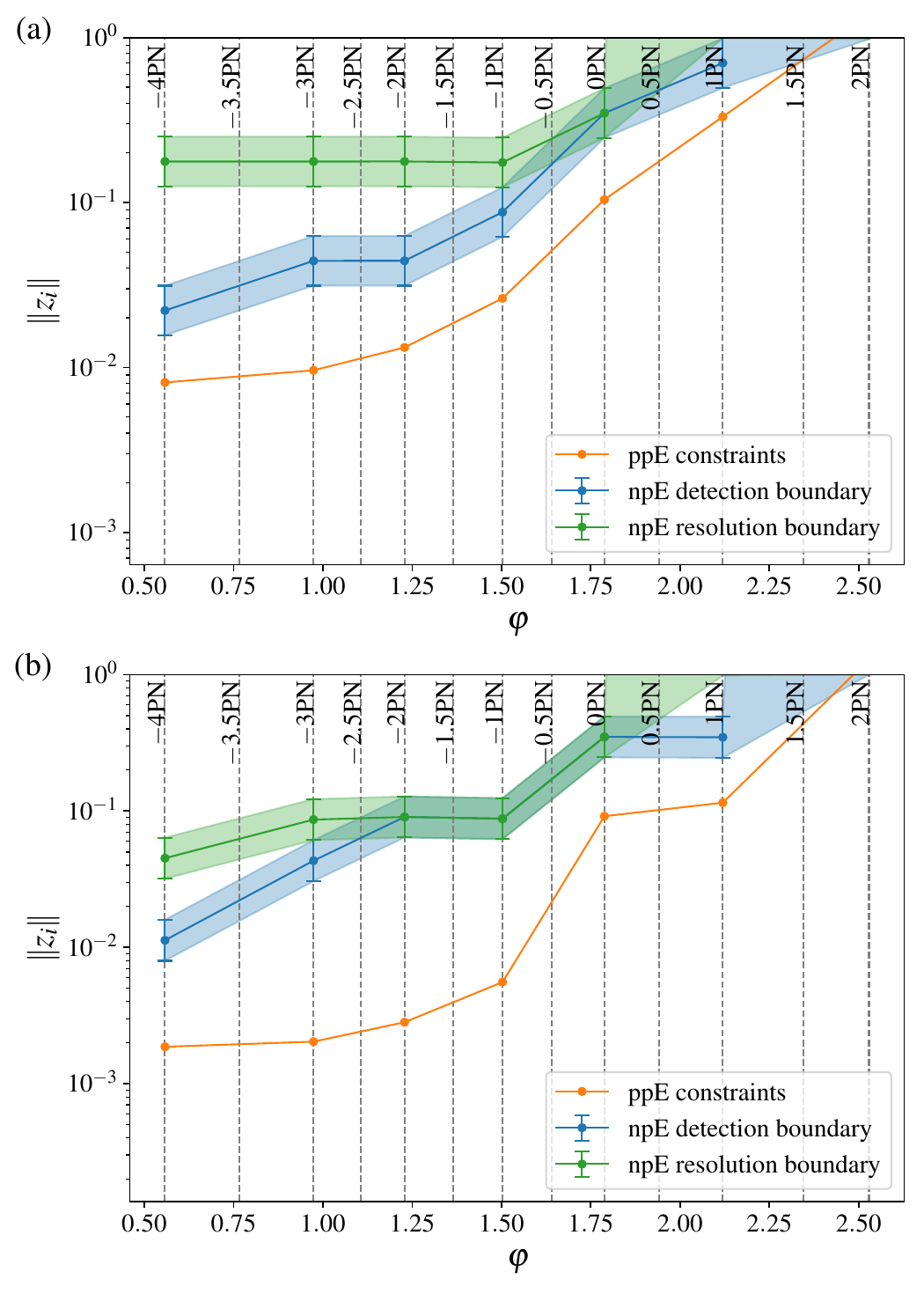}
    \caption{Sensitivity of the npE tests, for (a) the heavy BBH and (b) the light BBH. 
    The sensitivity is characterized by the detection boundary (minimal $\lVert z_i\rVert$ of injection with which the modification can be detected) in blue and the resolution boundary (minimal $\lVert z_i\rVert$ of injection with which the type of the modification can be resolved) in orange. 
    Each boundary is determined with a range of uncertainty, which is represented by a shade of the corresponding color. 
    The orange curve is the mapped ppE constraints when GR is injected. As explained in the main text, these ppE constraints are roughly the same as a detection boundary of the ppE test. 
    Observe that across the ppE-dense region of the latent space, the npE test is less sensitive than the ppE test in terms of detection. However, the npE test has the additional ability of resolving the type of modifications in one parameter estimation run.
    }
\label{fig:test_deviation_boundaries}
\end{figure}

The resulting boundary locations are summarized in Fig.~\ref{fig:test_deviation_boundaries}. For comparison, the ppE constraints from Sec.~\ref{sec:constrain_gr} are added to the plots. We have checked that the ppE 90\% CIs of $\beta_\mathrm{ppE}$ critically touch $0$ when the modifications are injected at the same level of these constraints. Therefore, we may think of the ppE constraints as an approximation to the ppE detection boundary. 
Due to the correlation across the npE latent space, the npE detection boundary is always above the ppE one, and the npE resolution boundary is always above the npE detection boundary (if not overlapping).
We note that, although the npE test is not as good as the ppE test at detecting a modification, it facilitates resolution of the theory type, which is impossible with a ppE test. 
Moreover, the boundaries in Fig.~\ref{fig:test_deviation_boundaries} are not constant limits as they should scale by $1/\mathrm{SNR}$. With that in mind, future observation of high-SNR events will make it possible to detect smaller non-GR deviations.

We note that the Bayesian parameter estimation runs above only involve odd $b_\mathrm{ppE}$'s in the npE training set. One may certainly add even $b_\mathrm{ppE}$'s to the training set, retrain the networks, and expect the augmented npE framework to detect these even-$b_\mathrm{ppE}$ modifications following the same trend as given in Fig.~\ref{fig:test_deviation_boundaries}.
However, the current training set does not necessarily limit the capability of the npE framework to detect and resolve these even-$b_\mathrm{ppE}$ modifications. 
In Appendix~\ref{apd:detect_ppe_untrained}, we show that our npE framework can properly resolve (detect) an injected $-1.5$PN ($0.5$PN) modification given that the magnitude of the injected modification is above the interpolated resolution (detection) boundary in Fig.~\ref{fig:test_deviation_boundaries}. 
This means that the npE generalization of the ppE framework is robust under our prescription, and the interpolation lines between those markers in Fig.~\ref{fig:test_deviation_boundaries} has the actual meaning of detecting and resolving modified gravity theories of the associated $\varphi$ angle.

\subsection{Detecting high-PN-order deviations from GR} \label{sec:detect_higher_pn}
In this subsection, let us consider injections with modifications expressed in a PN expansion. Compared to Sec.~\ref{sec:detect_ppe}, the modifications here not only adopt a ppE term as the leading PN order effect, but they also contain corrections from higher PN order terms. 
We will then study whether the npE framework can detect modifications to GR that are different from those in the npE training set. Because the npE latent space continuously extends the ppE parametrization, we expect these PN-like differences to be still captured by the npE framework. 

\begin{figure*}
    \centering
    \includegraphics[width=0.98\textwidth]{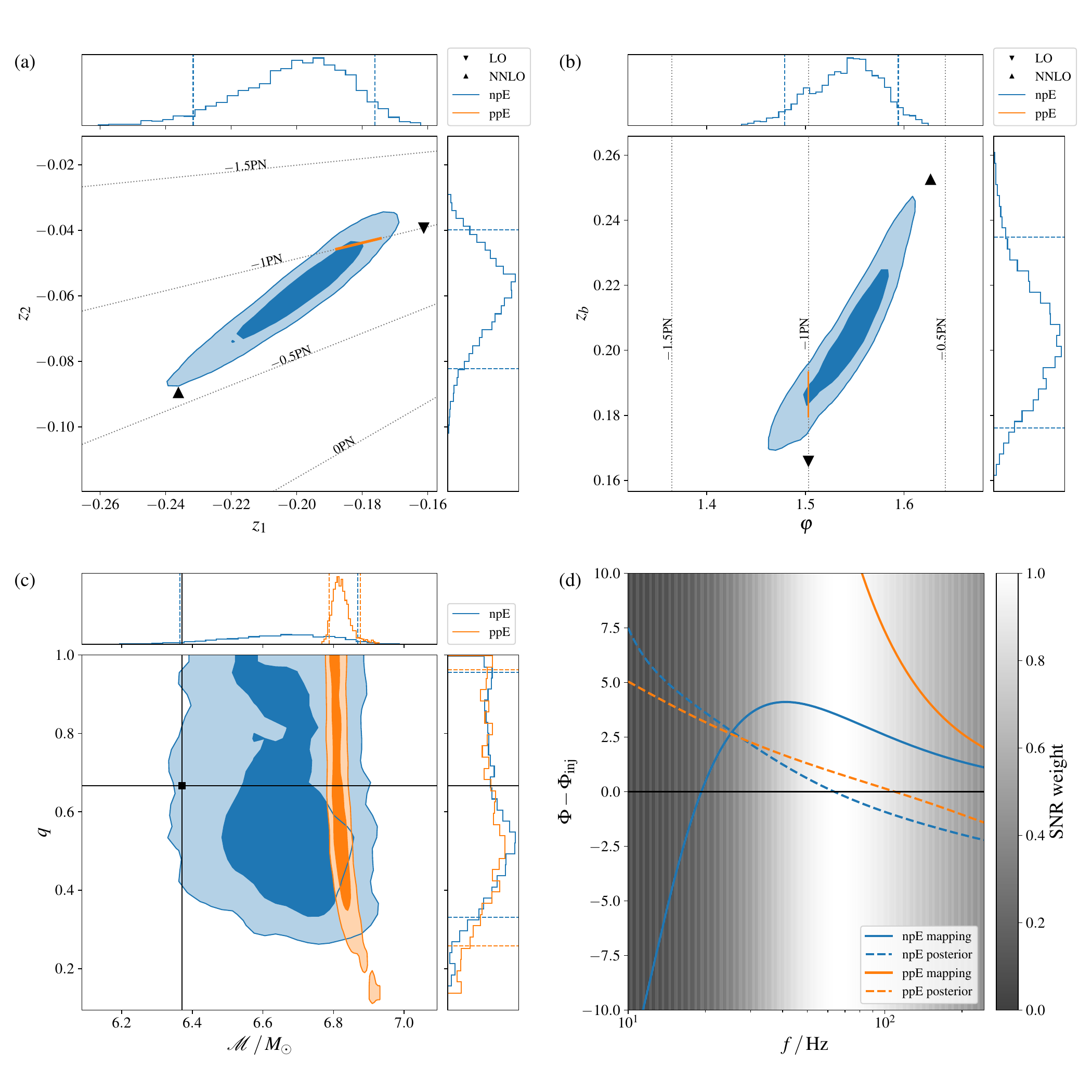}
    \caption{
    Recovering an EdGB modification with the npE framework. 
    The injected signal is generated with the light BBH, and the EdGB modification is computed to next-to-next-to-leading order with $\sqrt{\alpha_\mathrm{EdGB,inj}}=2.5\,\mathrm{km}$. Panels (a) and (b) show the posterior in the non-GR sector with the $z_1$-$z_2$ parametrization and the $z_b$-$\varphi$ parametrization, respectively. The up-pointing (down-pointing) triangle marks the npE representation of the EdGB phase modification calculated to next-to-next-to-leading (leading) order, using the npE encoder. Panel (c) shows the posterior of the mass parameters with the $z_1$-$z_2$ parametrization. The square dot at the intersection of solid black lines marks the injected values. 
    Other aspects of panels (a)--(c) follow the same format as in Fig.~\ref{fig:test_constrain_gr}.
    Panel (d) shows the reconstruction error of the total phase (GR plus modification) as a function of frequency. 
    The solid blue (orange) curve is the error of the npE (ppE) reconstruction taking source parameters as given in the injection, and non-GR parameters as given by the up-pointing (down-pointing) triangle in panels (a) and (b). The dashed blue (orange) curve is the error of the npE (ppE) reconstruction taking the median of the npE (ppE) marginalized posterior of each parameter, where the npE posterior is obtained with the $z_1$-$z_2$ parametrization.
    The background shading in panel (d) shows the weight of each frequency point in the SNR integral, proportional to $fA(f)^2/S_n(f)$.
    Observe that the npE posterior in the non-GR sector finds the theory type as slightly deviated from the leading $-1$PN order, and the npE recovery of the masses is less biased than the ppE one. Although the ppE estimate of the masses is more precise, it is also less accurate because it is completely biased, whereas our npE framework provides a much more accurate result, albeit it less precise.
    }
    \label{fig:test_edgb}
\end{figure*}

We take predictions from EdGB theory as an example, 
where the phase has been computed in a PN expansion up to the next-to-next-to-leading order~\cite{Shiralilou:2021mfl,Lyu:2022gdr}.
For the EdGB coupling strength, we choose $\sqrt{\alpha_\mathrm{EdGB,inj}}=2.5\,\mathrm{km}$, which is comparable to observational constraints from single BBH events~\cite{Nair:2019iur,Tahura:2018zuq,Yamada:2019zrb,Perkins:2021mhb}. 
We focus on the light BBH source for better resolution of the inspiral. We have checked that the EdGB correction never exceeds the small-coupling limit, when normalized with either component mass. 
We investigate the performance of the npE test in this particular setting, and summarize the results in Fig.~\ref{fig:test_edgb}.

Let us first take a look at how the npE framework models such a modification.
The npE representation of the given EdGB modification, found using the encoder mapping, is shown by the black up-pointing triangles in Fig.~\ref{fig:test_edgb} (a) and (b).  Observe that the npE representation is shifted toward $-0.5$PN order in response to the higher PN-order terms. This is to be compared to the leading-PN order, ppE counterpart, which is shown with down-pointing triangles on the $-1$PN line. 
Let us then use the npE up-pointing triangle to reconstruct the phase and compare this with the original EdGB injection. The difference, namely the modeling error, is shown by the blue solid curve in Fig.~\ref{fig:test_edgb} (d). In comparison, the ppE counterpart, obtained with the ppE down-pointing triangle, is shown by the orange solid curve. 
We note that the npE modeling error is significantly smaller than the ppE one in the entire frequency range, confirming that the npE parametrization captures the higher-PN-order corrections in the EdGB modification much better than the ppE model. 

The difference in modeling also causes some difference in the signal recovery. In Fig.~\ref{fig:test_edgb} (a) and (b), the npE posterior spans a range of angles, extending from roughly $-1$PN to roughly $-0.5$PN. On the contrary, the ppE posterior has to stick to the $-1$PN line, given its fixed ppE index.
Interestingly, although the ppE posterior seems to be farther away from the true theory under the latent representation, its recovery of the signal is as successful as the npE one. This is suggested by the dashed curves in Fig.~\ref{fig:test_edgb} (d), which are generated based on the medians of the npE and ppE posterior, respectively. At frequencies that weigh the most for the calculation of the total SNR, both dashed curves indicate small phase recovery error.
The reason for such inconsistency between modeling and recovery is revealed in Fig.~\ref{fig:test_edgb} (c), where we plot the posterior of the masses (in the chirp mass-mass ratio space). Observe that the ppE posterior of the chirp mass is significantly more biased, compensating the greater modeling error in the $\Phi_\mathrm{mod}$ sector of the full waveform. 
Thus, the less biased recovery of the source parameters is also an advantage of the npE framework. 

\subsection{Detecting non-PN-like deviations from GR} \label{sec:detect_hidden_sector}
Let us finally consider injections with modifications that cannot be described as $\Phi_\mathrm{ppE}$ plus higher-PN-order corrections, i.e.~that do not admit a polynomial representation. 
We thus wish to determine whether we can use the npE framework to detect non-GR modifications that are completely different from those in the npE training set, and which simply cannot be modeled with a simple ppE model. 

We take the dark-photon interactions theory~\cite{Alexander:2018qzg} as an example, in which the phase modification to the inspiral is given by
\begin{align}
    \Phi_\mathrm{DP} =\,& \frac{3}{128\eta}(\pi Mf)^{-5/3} \Bigg[ \frac{20\alpha_\mathrm{DP}}{3} F_3\left(\frac{M}{\lambda_\mathrm{DP}}(\pi Mf)^{-2/3}\right) \notag\\
    &- \frac{5\gamma_\mathrm{DP}}{84}(\pi Mf)^{-2/3} \Theta(\pi\lambda_\mathrm{DP} f-1) \Bigg], \label{eqn:test_phi_dp}
\end{align}
where $\Theta(\cdot)$ is the Heaviside step function, and
\begin{align}
    F_3(x) =\,& \left(\frac{180+180x+69x^2+16x^3+2x^4}{x^4}\right) e^{-x} \notag\\
    & + \frac{21\sqrt{\pi}}{2x^{5/2}} \mathrm{erf}(\sqrt{x}),
\end{align}
with $\mathrm{erf}(\cdot)$ the error function. 
This dark-photon modification is characterized by $\alpha_\mathrm{DP}$, $\gamma_\mathrm{DP}$ and $\lambda_\mathrm{DP}$, which are non-GR parameters. In particular, $\alpha_\mathrm{DP}$ and $\gamma_\mathrm{DP}$ are related to the distribution of dark matter in the source binary, and $\lambda_\mathrm{DP}$ is the Compton wavelength of the dark photon that mediates the interactions between those dark-matter particles.
Observe that the term proportional to the Heaviside function features a dipole activation at a frequency controlled by $\lambda_\mathrm{DP}$. 
On the other hand, when $({M}/{\lambda_\mathrm{DP}})(\pi Mf)^{-2/3}\gtrsim 1$, the $F_3$ term is nonvanishing and an attempt to PN expand this term will not result in convergent behavior~\cite{Alexander:2018qzg}.

Although Eq.~\eqref{eqn:test_phi_dp} is developed for binary neutron stars, we apply it to the BBHs that we study in this paper, as a proof of principle. 
Again, we focus on the light BBH source for better resolution of the inspiral. We assume the dark charge is distributed such that $\alpha_\mathrm{DP,inj}=0.05$ and $\gamma_\mathrm{DP,inj}=0.02$, which is within the range of validity of Eq.~\eqref{eqn:test_phi_dp}~\cite{Alexander:2018qzg}. 
For the dark photon, we consider two injected Compton wavelengths, $\lambda_\mathrm{DP,inj}=1400\,\mathrm{km}$ and $300\,\mathrm{km}$. 
In the first case, the dipole emission is activated at $64\,\mathrm{Hz}$. We have seen in Fig.~\ref{fig:test_edgb} (d) that this is roughly where most of the SNR is aggregated. 
In the second case, the dipole emission is activated at $273\,\mathrm{Hz}$ (or $0.020/M_\mathrm{inj}$), which is not within the frequency range of the injected signal; in this case, the dark-photon modification is proportional to the $F_3$ function in Eq.~\eqref{eqn:test_phi_dp}. 
Moreover, the $\lambda_\mathrm{DP,inj}$ value is chosen such that the argument of $F_3$ is slightly above unity around $64\,\mathrm{Hz}$. Thus, this scenario features nonperturbative behavior that is encoded by $F_3$.
We could further reduce $\lambda_\mathrm{DP,inj}$ to make $F_3$ even more nonperturbative, but doing so would also reduce the overall magnitude of $F_3$, which would eventually bring us back to a GR injection.

\begin{figure*}
    \centering
    \includegraphics[width=0.98\textwidth]{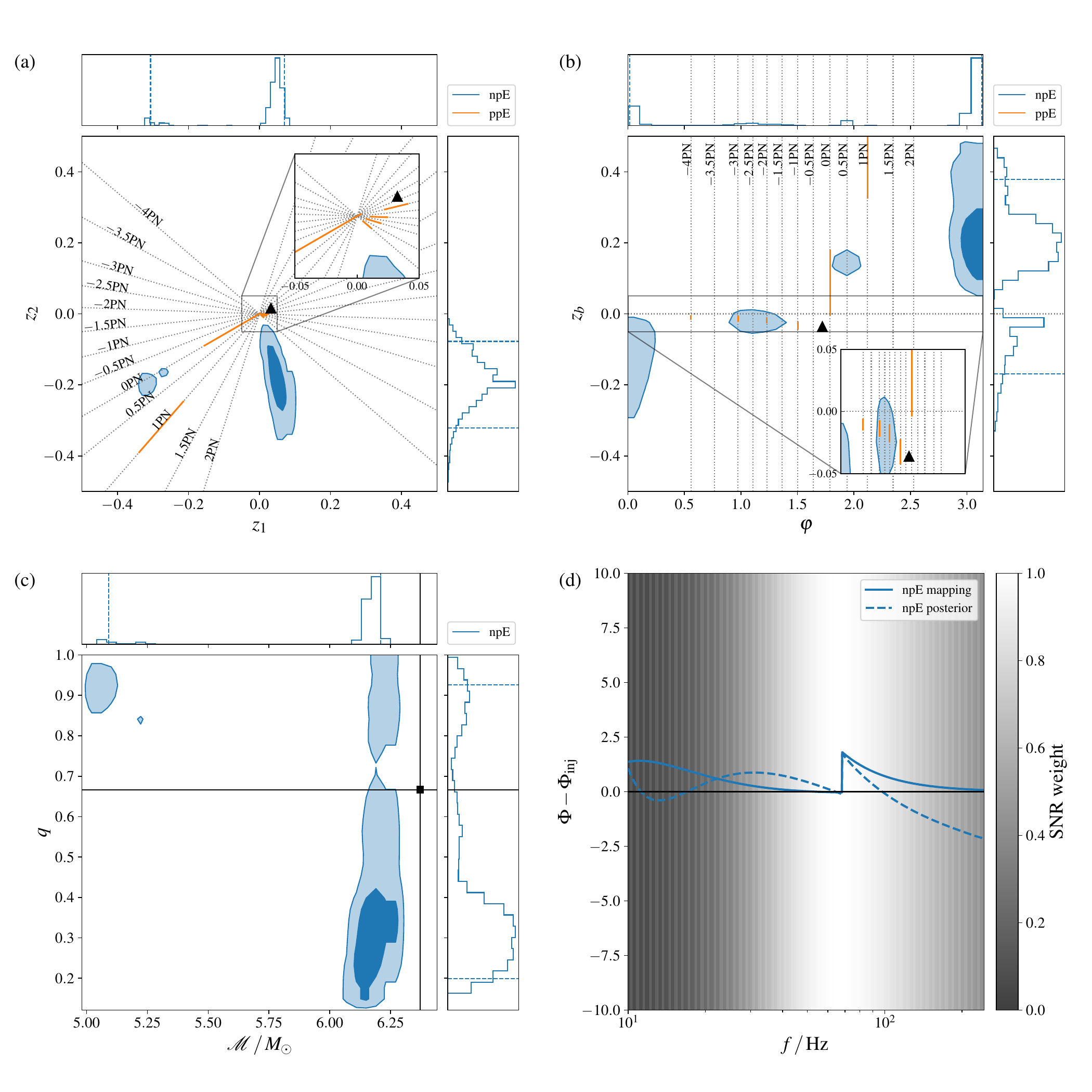}
    \caption{
    Using the npE framework to detect a dark-photon modification featured by a dipole activation. The injected signal is generated with the light BBH and $\lambda_\mathrm{inj}=1400\,\mathrm{km}$. 
    The npE results in panels (a)--(d) follow the same format as in Fig.~\ref{fig:test_edgb}. 
    For the ppE test, we show in panels (a) and (b) the 90\% credible region of the $\beta_\mathrm{ppE}$ posterior from a selection of $b_\mathrm{ppE}$'s, like those in Fig.~\ref{fig:test_constrain_gr}. 
    However, the insets of panels (a) and (b) now purely zoom in the central regions to show the ppE results better, and should not be interpreted as rerunning the npE test with a narrower prior range. 
    Observe that, although the modification is not covered by the npE parametrization, the majority of the npE posterior in the non-GR sector favors neither GR nor any of the ppE theories, indicating a non-PN-expandable modification in the signal.
    }
    \label{fig:test_darkphoton_dipole}
\end{figure*}

\begin{figure*}
    \centering
    \includegraphics[width=0.98\textwidth]{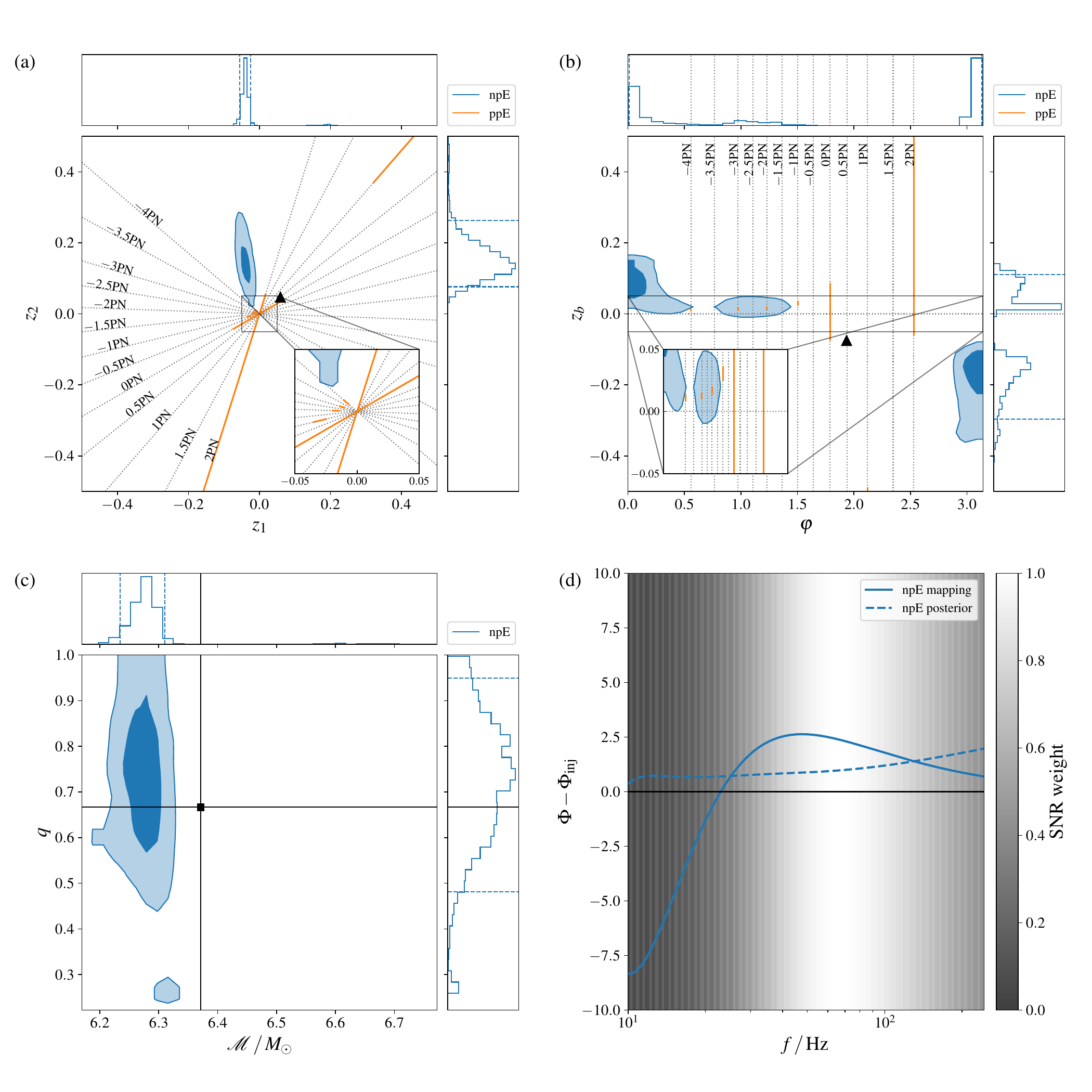}
    \caption{
    Using the npE framework to detect a dark-photon modification featured by the non-PN-expandable $F_3$ function. The injected signal is generated with the light BBH and $\lambda_\mathrm{inj}=300\,\mathrm{km}$.
    Panels (a)--(d) follow the same format as in Fig.~\ref{fig:test_darkphoton_dipole}.
    Similar to the observation in Fig.~\ref{fig:test_darkphoton_f3}, although the modification is not covered by the npE parametrization, the majority of the npE posterior in the GR sector favors neither GR nor any of the ppE theories, indicating a non-PN-expandable modification in the signal. 
    }
    \label{fig:test_darkphoton_f3}
\end{figure*}

Figures~\ref{fig:test_darkphoton_dipole} and \ref{fig:test_darkphoton_f3} show the npE results of the above two injection cases, respectively and allow us to make several observations.  
First, observe that for both injections, the npE posteriors recover $z_i$ very differently from the npE representation of the injected $\Phi_\mathrm{DP}$ (the black up-pointing triangles). 
For the first injection, Fig.~\ref{fig:test_darkphoton_dipole} (d) indicates that the npE latent space does not cover anything like a dipole activation, as the step function in Eq.~\eqref{eqn:test_phi_dp} is not eliminated in errors reconstructed from either the representation point or the posterior median. 
For the second injection, Fig.~\ref{fig:test_darkphoton_f3} (d) suggests that the npE model covers the nonperturbative $F_3$ by the same criteria in Sec.~\ref{sec:detect_higher_pn}, as the shape and magnitude of the phase reconstruction errors resemble those in Fig.~\ref{fig:test_edgb} (d) for EdGB.
However, this is incorrect. 
For the injection, the npE representation point lies right on the $0.5$PN line, so if ppE had been considered and all indices had been explored, the same reconstruction could be reproduced. 
For the recovery, the npE posterior differs a lot from the representation point, with a majority disfavoring all ppE theories included by the latent space. Moreover, the posterior recovery of the chirp mass fails, as shown in Figs.~\ref{fig:test_darkphoton_dipole} (c) and \ref{fig:test_darkphoton_f3} (c), leading to a strong bias in chirp mass. 
All these observations suggest that the npE framework cannot model those non-PN-expandable modified signals more correctly than what the ppE framework can achieve.

Despite the above conclusions, an npE analysis has very powerful and important advantages over a ppE analysis. First, for both injections, the 90\% CC in the $z_1$-$z_2$ plane excludes the GR region at the center. This means that the npE analysis can signal that the injection is not consistent with GR. Second, the majority of the $z_1$-$z_2$ and $z_b$-$\varphi$ posterior disfavors all ppE theories included by the latent space. 
Therefore, with the npE framework, not only can we still detect the modification to GR, but we can even say that it did not admit a PN expansion.
Third, these conclusions can be arrived at by carrying out parameter estimation only once for each signal injected.
In a ppE analysis, however, one would have to repeat the parameter estimation for every possible $b_\mathrm{ppE}$ [see results from example runs in panels (a) and (b) of Figs.~\ref{fig:test_darkphoton_dipole} and \ref{fig:test_darkphoton_f3} with odd $b_\mathrm{ppE}$'s between $-13$ and $-1$]. Although some of these ppE subtests can alert of a deviation from GR, they always suggest a PN-type modification by construction and miss the fact that the actual modification is not PN-expandable. 
We thus conclude that, even though the npE framework is not perfect, it is still preferred over the ppE framework, since it is more general and computationally efficient.  

\section{Conclusions} \label{sec:conclusions}
In this work, we introduced the npE framework for efficient theory-agnostic GW tests of GR, using a deep-learning parametrization of non-GR theories extended from the ppE framework. 
The key component of the npE framework is a VAE. 
When trained with waveforms from a list of ppE theories, the VAE encoder finds a continuous representation for these waveforms in a dimensionally-reduced latent space. Then, the VAE decoder reconstructs similar waveforms using points in the latent space as parameters. 
For GW test of GR, the decoder can be used as a waveform model, with the latent space providing the non-GR parameters.
Because the latent space unifies many non-GR theories that fit within the ppE framework, one only needs to run parameter estimation once to test all of them, which makes the npE framework more efficient than the ppE one.
Moreover, because the gaps between ppE theories in the latent space are filled continuously with non-ppE modifications, the npE test of GR is more theory-agnostic. 

In order to test the npE framework and reach the above conclusions, we focused on a parametrized modification to the inspiral GW phase of the IMRPhenomD waveform, and created a modified gravity dataset with $\Phi_\mathrm{ppE}$ in a variety of ppE theories and from a population of BBHs. 
We equipped the npE framework with a two-dimensional latent space $z_{1,2}$, in which GR sits at the center ($z_i=0$) and the npE phase modification $\Phi_\mathrm{npE}$ is proportional to the polar radius $\lVert z_i\rVert$. 
We decomposed $\Phi_\mathrm{npE}$ as the product of $\lVert z_i\rVert$, a shape function $S$, and a scale function $T$. 
We trained the VAE network to find a representation for $S$ with the polar angle in the latent space. After training, we found that ppE theories map to radial lines that are separated and ordered in the latent space. Gaps between those ppE lines continuously extend the ppE parametrization. 
We then trained a secondary network to emulate $T$, so that $\Phi_\mathrm{npE}$ maps the unit circle $\lVert z_i\rVert=1$ to modifications that are comparable with GR terms, regardless of the BBH source properties. 
For parametrized tests of GR with GW data, we applied $\Phi_\mathrm{npE}$ as a phase modification to the IMRPhenomD model, and ran Bayesian parameter estimation to recover the latent parameters $z_i$. The prior for $z_i$ was chosen to be flat within $\lVert z_i\rVert=1$, which we referred to as the EFT prior boundary. 
A deviation from GR is then detected if the $z_i$ posterior excludes $z_i=0$, and this posterior then can identify and resolve the type of non-GR modification that is best supported by the data in the latent space. 
We note that our VAE is customized with additional features added to the vanilla formulation, including the separation of the latent angle and radius, the symmetrization of the latent representation, and the pseudo-PN expansion embedded in the decoder to better fit the physical mission.

To demonstrate our prescription, we performed Bayesian parameter estimation on simulated BBH GW signals observed by a detector network with O5 designed sensitivity, and studied the npE recovery of potential non-GR modifications. 
We verified that the npE framework can be used to detect non-GR modifications in the GW signal and resolve the type of non-GR theory that is encoded in the data, provided the non-GR modification is large enough. 
Although the npE framework cannot dig as deep as the ppE framework to detect non-GR modifications (due to correlation between theories across the latent space), the npE framework can not only detect modifications, it can also classify the type of deviation detected and it can do so with a single parameter estimation run. The same task in the ppE framewrk would require many runs for many different ppE indices.

The fact that npE tests of GR are more theory-agnostic than ppE tests is particularly showcased with signals simulated in EdGB gravity with higher PN order correction and signals in a theory with dark-photon interactions. 
In the first case, the EdGB modification to the signal can be detected by the npE framework, with the npE $z_i$ posterior correctly recovering a trend that suggests higher-PN order corrections in the signal. Moreover, the mass recovery using $\Phi_\mathrm{npE}$ as the model also returns significantly less bias in comparison to recovery with $\Phi_\mathrm{ppE}$.
In the second case, the non-PN dark-photon modifications can still be detected with the npE framework with the npE $z_i$ posteriors selecting a region of the latent space far from the GR region. 
These observations suggest that the npE tests of GR can handle theories beyond the ppE description in a reasonable way.

Our results have various direct applications to GW tests of GR in the upcoming LVK O5 run, and perhaps, even for O1-O4 observations. 
The efficiency of npE tests will be especially desirable, once the number of signals increases as expected in O5 and with next-generation detectors. 
For signals detected at potentially higher SNRs, the npE framework offers less biased recovery, assuming that GR is tested against modifications that deviate slightly from the standard ppE form. 
In addition, the npE posterior can also raise an alarm when an observed non-GR effect cannot be modeled by a PN expansion.

The npE framework uses the VAE and the latent space to address both the need of being computationally efficient and the need of being theory-agnostic, while previous work, like~\cite{Perkins:2022fhr,Wolfe:2022nkv}, focused on either one or the other problem. 
In particular, in~\cite{Wolfe:2022nkv}, the efficiency of a ppE-like test of GR was improved by a hybridized sampling method that reduces the number of likelihood evaluations. This procedure starts with a parameter estimation run assuming GR, which is then followed by a set of cheaper estimation runs with ppE-like modifications included. 
In comparison, the npE test runs parameter estimation only once with GR plus only two npE latent parameters. 
The number of likelihood evaluations of both schemes should be comparable to that of a standard GR parameter estimation run. 
With that being said, one could think that the npE test could be slowed down in each likelihood evaluation step, due to the use of neural networks in the waveform model. However, in this work, we observed no significant slow-down of waveform evaluation using the npE template. 
Therefore, we expect our prescription to not cost significantly more computational time in comparison to that proposed in~\cite{Wolfe:2022nkv}.
Moreover, the computational procedure of the npE test is easier to manage as there will be only one parameter estimation run involved. 

In~\cite{Perkins:2022fhr}, the authors worked on an upgraded ppE framework with higher-PN-order terms, and proposed a way to regularize the prior for those higher-PN-order coefficients in order to obtain informative posteriors. This results in a more theory-agnostic version of the ppE framework that can be used for practical purposes. 
In comparison, our npE framework can also recover signals with higher-PN-order corrections at gaps between ppE lines in the latent space, and the EFT prior boundary serves similarly as the regularization on the prior. 
Apart from that, there are also regions in the latent space that are far from any ppE lines, allowing the npE test to raise an alarm when the GW signal contains a non-GR effect that cannot be modeled by any PN expansion. This features the flexibility of the neural networks, which is new in the npE framework. 

Our npE framework is built on a training set composed of ppE modifications to the GW phase of the inspiral of BBHs. 
This training set can be augmented to suit the npE framework for more general purposes. For example, NSs can be included in the training set, provided that $\bar{f}_\mathrm{min}$ of the frequency grid is reduced accordingly to cover $f_\mathrm{low}$ of the detector with NS masses. 
One may also teach the VAE to learn beyond-ppE theories or modifications to the merger and the ringdown, possibly with amplitude modifications included as well. Doing so may require the shape model to be redesigned, and the dimension of the latent space to be increased. 
With additional effects properly modeled beyond the inspiral phase, the boundaries in Fig.~\ref{fig:test_deviation_boundaries} can be lowered for detection of smaller non-GR deviations.

The npE framework we developed here was implemented on an IMRPhenomD model for simplicity, but in principle, the same idea can be applied to any base GR model, especially those including precession and higher harmonics that may appear more frequently in future observations. The performance of the npE framework was demonstrated with injection-recovery runs in which the signals were not injected in any noise realization, and the results were presented as the averaged outcome. The actual impact from specific noise realizations (like those in the real GW data) may be of interest for future studies.

In our current prescription, the npE test adopts an EFT prior boundary, at which the modifications are comparable to the corresponding effects in GR. 
If an alternative boundary for the current most stringent constraints is wanted, one may rescale the training set and retrain the secondary network. This also applies to updates to the boundary for any other reason.
On the other hand, further studies may benefit from fixing the prior for multiple events across a certain observation period, so that the npE posterior of each single event can be stacked to further improve resolution (and hierarchical inference~\cite{Isi:2019asy} still applies).
In this way, the npE framework opens the door to physics-informed machine-learning in the development of GW tests of GR for current and future ground- and space-based detectors.

\section{Acknowledgments}
Y.~X. and N.~Y. acknowledge support from the Simons Foundation through Award No.~896696, the National Science Foundation (NSF) Grant No.~PHY-2207650, and the National Aeronatucis and Space Agency (NASA) Award No.~80NSSC22K0806.
Y.~X. also acknowledges support from the Illinois Center for Advanced Studies
of the Universe (ICASU)/Physics Graduate Fellowship.
D.~C. acknowledges support from NSF Grant No.~PHY-2117997.
G.~N. acknowledges NSF support from AST-2206195, and a CAREER grant, supported in-part by funding from Charles Simonyi.
This work used the Delta advanced computing and data resource which is supported by NSF (award OAC~2005572) and the State of Illinois. Delta is a joint effort of the University of Illinois Urbana-Champaign (UIUC) and its National Center for Supercomputing Applications (NCSA).
This work also made use of the Illinois Campus Cluster, a computing resource that is operated by the Illinois Campus Cluster Program (ICCP) in conjunction
with NCSA, and is supported by funds from the UIUC.
The authors thank Elise S\"{a}nger and Soumen Roy for reviewing the document and providing helpful feedback. This document is given the LIGO DCC No.~P2400078.

\appendix
\section{Computational Cost of the npE Template} \label{apd:npe_time_complexity}

In Sec.~\ref{sec:train}, one may be concerned about the evaluation time of the npE template, which one could imagine could be significantly long for a sufficiently complicated neural-network implementation. 
With the computational settings described in Sec.~\ref{sec:test_setup}, we tested the evaluation of the npE template (IMRPhenomD plus npE phase modification) versus the ppE template (IMRPhenomD plus ppE phase modification), on a single core of an Intel Xeon Platinum 8358 CPU. 
For the heavy source, the npE evaluation time is about $\tau_\mathrm{npE}=3.3\,\mathrm{ms}$, while the ppE evaluation takes about $\tau_\mathrm{ppE}=1.5\,\mathrm{ms}$. For the light source, the results are $\tau_\mathrm{npE}=14.6\,\mathrm{ms}$ and $\tau_\mathrm{ppE}=11.7\,\mathrm{ms}$.
Thus, we conclude that for the cases studied here, the npE implementation at most doubles the evaluation time of the template.

The mass of the BBH source impacts the waveform evaluation time through the length of the signal: lighter sources have longer signals, and hence it takes a longer time to evaluate the corresponding waveforms. 
In the case of the ppE template, the time complexity scales almost linearly with the signal length. 
In comparison, the additional time complexity of the npE template resides mostly in the forward propagation of $\mathcal{D_U}$, $\mathcal{D_V}$ and $\mathcal{T}$, which is rather constant (observe that the difference between $\tau_\mathrm{npE}$ and $\tau_\mathrm{ppE}$ is not much affected by the source).
Such an npE overhead is only significant for the heavy source because the signal is short and the non-neural part of the template is fast.

\section{NpE Reconstruction of Phase Modifications in Specific ppE Theories} \label{apd:npe_reconstruction}

In this appendix, we check whether the npE framework trained in Sec.~\ref{sec:train} reproduces the ppE modeling of non-GR theories.
We consider EdGB gravity and dCS gravity as two examples, compute their phase modification in the ppE framework, and use the npE framework to reconstruct the phase modification. 
In the EdGB case, we consider the same settings as in Sec.~\ref{sec:detect_higher_pn}, i.e.~we assume a nonspinning BBH source with component masses $m_1=9\,\mathrm{M_\odot}$ and $m_2=6\,\mathrm{M_\odot}$, and take the EdGB coupling to be $\sqrt{\alpha_\mathrm{EdGB}}=2.5\,\mathrm{km}$, except that here we only compute the phase modification to leading PN order. 
In the dCS case, we again choose a BBH source with $m_1=9\,\mathrm{M_\odot}$ and $m_2=6\,\mathrm{M_\odot}$, and take the dCS coupling to be$\sqrt{\alpha_\mathrm{dCS}}=8.5\,\mathrm{km}$, which saturates the $90\%$ constraint using mass and equatorial plane measurements of an isolated neutron star~\cite{Silva:2020acr}. The leading PN order dCS phase modification is computed following~\cite{Yunes:2016jcc}. We note that the resulting dCS phase modification vanishes for nonspinning BBHs, and therefore, in this case, we consider the source to have anti-aligned spins $\chi_1=-0.5$ and $\chi_2=0.5$.

\begin{figure}
    \centering
    \includegraphics[width=0.48\textwidth]{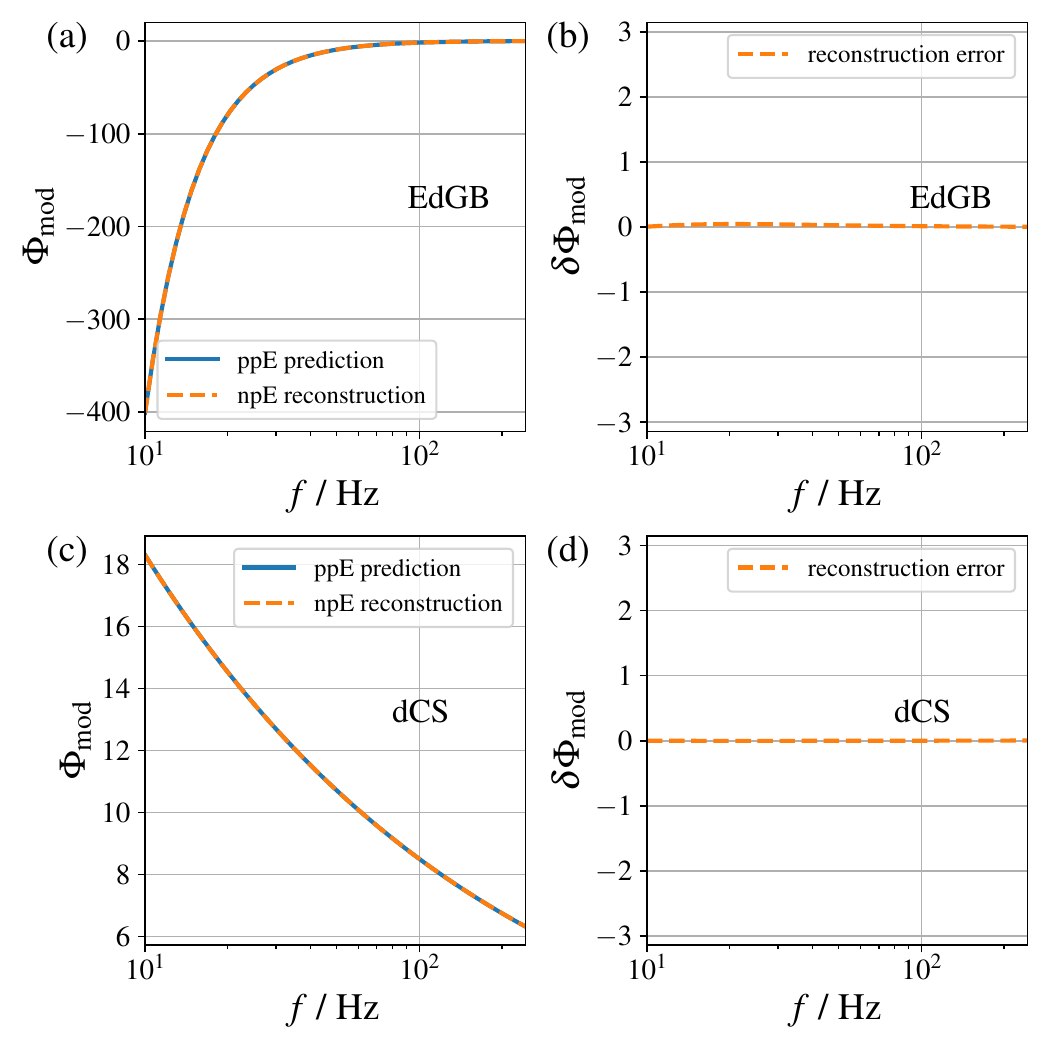}
    \caption{NpE reconstruction of ppE phase modifications. Panels (a) and (b) [(c) and (d)] show an example in the EdGB (dCS) gravity. In the EdGB case, the source is a nonspinning BBH with $m_1=9\,\mathrm{M_\odot}$ and $m_2=6\,\mathrm{M_\odot}$, and the EdGB coupling is $\sqrt{\alpha_\mathrm{EdGB}}=2.5\,\mathrm{km}$. In the dCS case, the source is a BBH with $m_1=9\,\mathrm{M_\odot}$, $m_2=6\,\mathrm{M_\odot}$ and aligned spins $\chi_1=-0.5$, $\chi_2=0.5$. The dCS coupling is $\sqrt{\alpha_\mathrm{dCS}}=8.5\,\mathrm{km}$. 
    In panels (a) and (c), we show the ppE (leading PN order) prediction of the phase modification and its npE reconstruction, in the frequency range from $10\,\mathrm{Hz}$ to the inspiral cutoff $Mf=0.018$. Panels (b) and (d) further shows the difference between the ppE and npE results, namely the npE reconstruction error, in panels (a) and (c), respectively. The $y$-axis in panels (b) and (d) ranges from $-\pi$ to $\pi$, and we observe that the npE reconstruction error is $\ll\pi$, suggesting that the npE framework can accurately reproduce the ppE modeling of these two example theories.}
    \label{fig:npe_reconstruction}
\end{figure}

The npE reconstruction of the phase modifications follows the same procedure as described in Sec.~\ref{sec:detect_higher_pn}. That is, given the ppE (leading PN order) prediction of the EdGB/dCS phase modification, we first find its npE representation in the latent space using the encoder mean, and then we compute $\Phi_\mathrm{npE}$ at this representation point. We compare the ppE prediction to the npE reconstruction, and show the results in Fig.~\ref{fig:npe_reconstruction}. 
Observe that the npE reconstruction error is always $\ll\pi$ despite the fact that both examples are as large as possible, since we have saturated the current observational constraints. 
This means that our npE framework has been well-trained to accurately reproduce the ppE modeling of non-GR theories.

\section{Reference Angle for the $z_b$-$\varphi$ Parametrization} \label{apd:angle_ref}
\begin{figure}
    \centering
    \includegraphics[width=0.48\textwidth]{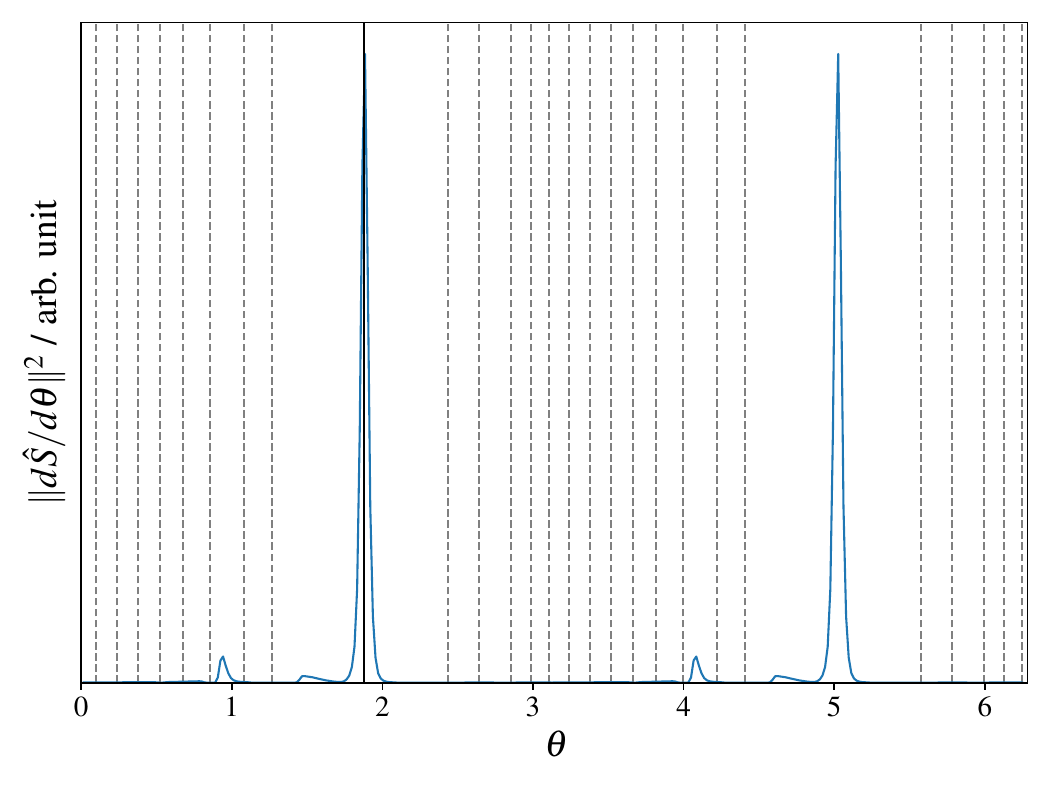}
    \caption{Variation of the reference angle, in order to locate the reference angle for the $z_b$-$\varphi$ parametrization. The blue curve shows the rate of shape variation with polar angle. Observe that the rate of variation presents two clear peaks, each taking place in one of the two ppE-sparse regions. The first peak, marked by the black vertical line at $\theta_\mathrm{ref}=1.88$, is the one used in this paper to define the reference angle. }
    \label{fig:angle_ref}
\end{figure}

In Sec.~\ref{sec:train}, we defined the $z_b$-$\varphi$ parametrization based on a reference polar angle $\theta_\mathrm{ref}$ in the latent space. This reference angle should be placed in a ppE-sparse region, and should be related to the place where the shape reconstruction varies the most rapidly. Here, we define
\begin{align}
    \theta_\mathrm{ref}=\mathrm{arg}\max_{0\leq\theta<\pi} \bigg\lVert\frac{d\hat{S}}{d\theta}\bigg\rVert^2. \label{eqn:angle_ref}
\end{align}
The maximum is taken only for the angle range $[0,~\pi)$ because the latent representation is designed to be symmetric. 
In Fig.~\ref{fig:angle_ref}, we visualize the magnitude of the derivative in Eq.~\eqref{eqn:angle_ref}. The maximum can be easily located at the two peaks, and, as expected, these peaks occur in the gaps between the ppE-dense regions. 
The first peak, within the range of $[0,~\pi)$, is where we define the reference angle, and we find
\begin{align}
    \theta_\mathrm{ref}=1.88.
\end{align}

\section{Detecting Leading-PN-Order Deviations not Covered by the npE Training Set} \label{apd:detect_ppe_untrained}
In Sec.~\ref{sec:detect_ppe}, we investigated the use of the npE framework for detecting a modification of the form $\Phi_\mathrm{ppE}$, i.e.~containing only the leading-PN-order term. The results shown in Figs.~\ref{fig:test_deviation_examples} and \ref{fig:test_deviation_boundaries} cover the odd $b_\mathrm{ppE}$'s provided by the npE training set. 
For even $b_\mathrm{ppE}$'s, one may certainly add them to the training set, retrain the networks, and expect the augmented npE framework to detect these even-$b_\mathrm{ppE}$ modifications following the same trend as given in Fig.~\ref{fig:test_deviation_boundaries}. 
Here, however, let us try to detect these even-$b_\mathrm{ppE}$ modifications without augmenting our npE framework. In this way, we can investigate how robustly our npE framework generalizes the ppE parametrization. 

\begin{figure*}[htbp]
    \centering
    \includegraphics[width=0.98\textwidth]{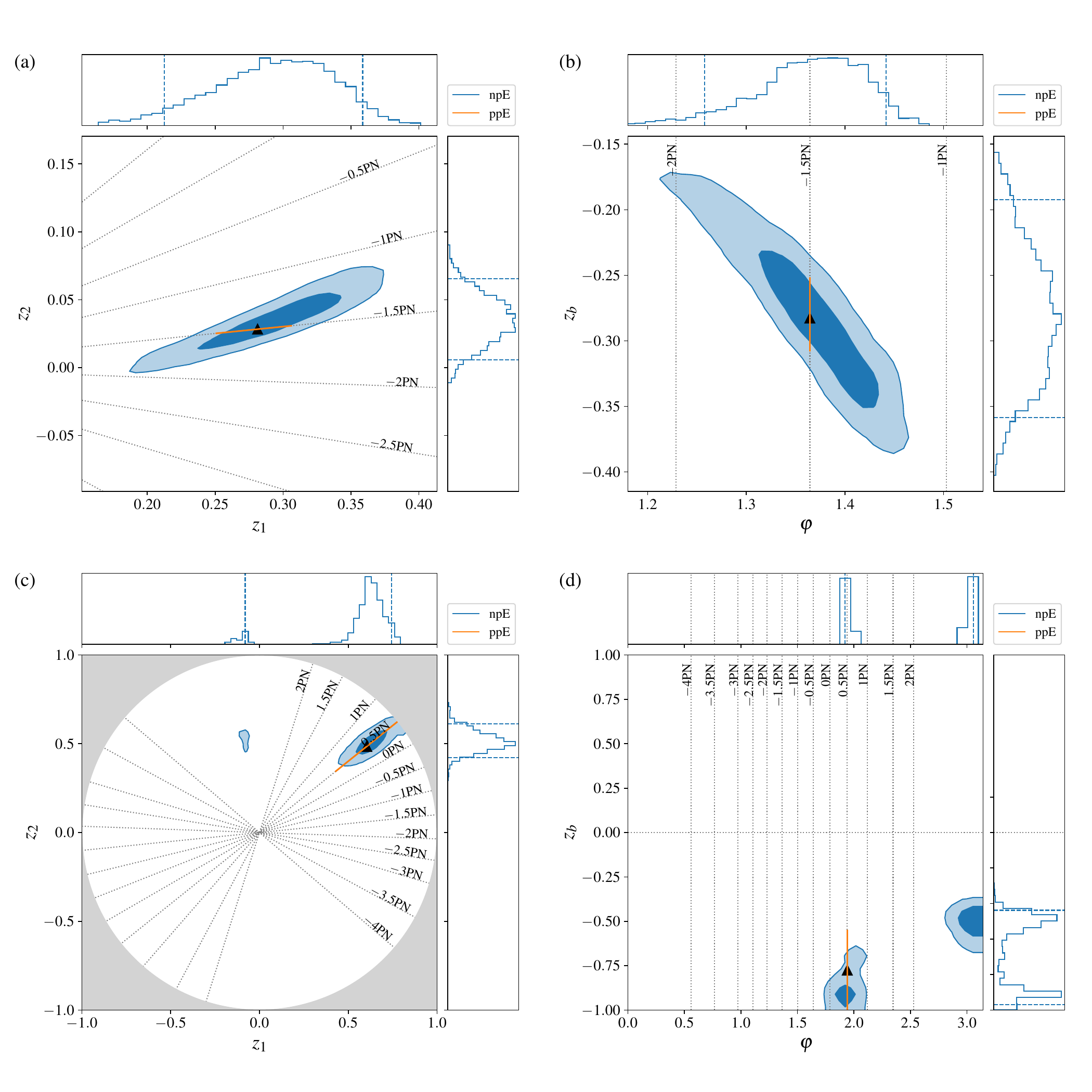}
    \caption{
    Posteriors recovering ppE modifications not included by the npE training set. The injected signals are generated with the heavy BBH source. 
    The modification chooses from two cases. One is a $-1.5$PN-order term with $\beta_\mathrm{ppE,inj}/\beta_\mathrm{ppE,max}=25\%$ [panels (a) and (b)], which is right above the interpolated npE resolution boundary in Fig~\ref{fig:test_deviation_examples} (a). 
    The other is a $0.5$PN-order term with $\beta_\mathrm{ppE,inj}/\beta_\mathrm{ppE,max}=100\%$ [panels (c) and (d)], which is above the interpolated npE detection boundary, but not necessarily above the resolution boundary in Fig~\ref{fig:test_deviation_examples} (a). 
    Results are obtained with both the $z_1$-$z_2$ parametrization [panels (a) and (c)] and the $z_b$-$\varphi$ parametrization [panels (b) and (d)].
    The plots follow the same format as in Fig.~\ref{fig:test_constrain_gr}.
    Observe that the npE recovery resolves the injected $-1.5$PN modification and detects the injected $0.5$PN modification, as expected. 
    }
    \label{fig:test_deviation_examples_untrained}
\end{figure*}

Figure~\ref{fig:test_deviation_examples_untrained} shows examples using the npE framework to recover injections with $-1.5$PN ($b_\mathrm{ppE,inj}=-8$) and $0.5$PN ($b_\mathrm{ppE,inj}=-4$) modifications. These choices allow us to cover modifications at both negative and positive PN orders. 
For each case, we choose $\beta_\mathrm{ppE,inj}$ in reference to the boundaries shown in Fig.~\ref{fig:test_deviation_boundaries} and check whether a detection or resolution is achieved as expected. 
For the $-1.5$PN modification, we choose $\beta_\mathrm{ppE,inj}$ to be right above the interpolated resolution boundary, and we confirm that the resolution is achieved as the $z_b$ posterior excludes zero, and the $\varphi$ posterior excludes both neighboring ppE lines. 
For the $0.5$PN modification, the resolution boundary is undetermined by Fig.~\ref{fig:test_deviation_boundaries}, so we choose $\beta_\mathrm{ppE,inj}/\beta_\mathrm{ppE,max}=100\%$ at maximum, which is at least above the interpolated detection boundary. In this case, we find that detection is indeed successful, as the $z_b$ posterior excludes zero. Although a resolution of the theory type is not expected, the npE posterior at least covers the $0.5$PN line in both the $z_1$-$z_2$ parametrization and the $z_b$-$\varphi$ parametrization. 
These observations suggest that the npE framework performs well when recovering those ppE modifications with $b_\mathrm{ppE}$'s not included in the training set, which proves the robustness of our npE prescription for generalizing the ppE framework. 

\bibliography{references}

\end{document}